\documentclass[preprint,aps,preprintnumbers,amsmath,amssymb]{revtex4}

\usepackage{epsfig}
\usepackage{graphicx}% Include figure files
\usepackage{dcolumn}% Align table columns on decimal point
\usepackage{bm}% bold math

\let\jnfont=\rm
\def\NPB#1,{{\jnfont Nucl.\ Phys.\ B }{\bf #1},}
\def\PLB#1,{{\jnfont Phys.\ Lett.\ B }{\bf #1},}
\def\EPJC#1,{{\jnfont Eur.\ Phys.\ Jour.\ C }{\bf #1},}
\def\PRD#1,{{\jnfont Phys.\ Rev.\ D }{\bf #1},}
\def\PRL#1,{{\jnfont Phys.\ Rev.\ Lett.\ }{\bf #1},}
\def\MPLA#1,{{\jnfont Mod.\ Phys.\ Lett.\ A }{\bf #1},}
\def\JPG#1,{{\jnfont J.\ Phys.\ G}{\bf #1},}
\def\CTP#1,{{\jnfont Commun.\ Theor.\ Phys.\ }{\bf #1},}
\def\JHEP#1,{{\jnfont JHEP \ }{\bf #1},}
\def\NPPS#1,{{\jnfont Nucl.\ Phys.\ Proc.\ Suppl.\ }{\bf #1},}

\def\oversim#1#2{\lower0.5ex\vbox{\baselineskip0pt\lineskip0pt
  \lineskiplimit0pt\everycr{}\tabskip0pt
  \halign{$\mathsurround0pt #1\hfil##\hfil$\crcr #2\crcr\sim\crcr}}}
\begin{document}

\preprint{\parbox{1.2in}{\noindent arXive:0810.0751}}

\title{
                  Anomaly of $Zb\bar b$ coupling revisited in MSSM and NMSSM}

\author{\ \\[2mm]  Junjie Cao$^1$,  Jin Min Yang$^2$ }

\affiliation{$^1$ Ottawa-Carleton Institute for Physics, Department
                  of Physics, Carleton University, Ottawa, Canada K1S 5B6 \\
$^2$Institute of Theoretical Physics and Kavli Institute for Theoretical Physics China,
Academia Sinica, Beijing 100190, China
     \vspace*{1.5cm}}

\begin{abstract}
The $Z b \bar{b}$ coupling determined from the $Z$-pole
measurements at LEP/SLD shows an about $3\sigma$ deviation from the
SM prediction, which would signal the presence of new physics in
association with the $Zb\bar b$ coupling. In this work we give a
comprehensive study for the full one-loop supersymmetric effects on
the $Z b \bar{b}$ coupling in both the MSSM and the NMSSM by
considering all current constraints which are from the precision
electroweak measurements, the direct search for sparticles and Higgs
bosons, the stability of Higgs potential, the dark matter relic
density, and the muon $g-2$ measurement. We analyze the characters
of each type of the corrections and search for the SUSY parameter
regions where the corrections could be sizable. We find that the
sizable corrections may come from the Higgs sector with light $m_A$
and large $\tan \beta$, which can reach $-2\%$ and $-6\%$ for
$\rho_b $ and $\sin^2 \theta_{eff}^b$, respectively. However,
such sizable  negative corrections are just opposite to
what needed to solve the anomaly. We also scan over the allowed
parameter space and investigate to what extent supersymmetry can
narrow the discrepancy. We find that under all current constraints,
the supersymmetric effects are quite restrained and cannot
significantly ameliorate the anomaly of $Zb\bar b$ coupling.
Compared with $\chi^2/dof = 9.62/2$ in the SM,  the MSSM and NMSSM
can only improve it to $\chi^2/dof = 8.77/2$ in the allowed
parameter space. Therefore, if the anomaly of $Zb\bar b$ coupling
is not a statistical or systematic problem, it would suggest new
physics beyond the MSSM or NMSSM.

\vspace*{1cm}
\end{abstract}


\maketitle

\section{introduction}

Although most of the electroweak data are consistent with the
Standard Model (SM) to a remarkable precision, there are still some
experimental results difficult to accommodate in the SM framework.
A well known example is that the effective electroweak
mixing angle  $\sin^2 \theta_{eff}$ determined from
the leptonic asymmetry measurements is much lower than
the value determined from the hadronic asymmetry measurements \cite{2005ema,Grunewald},
and the averaged value over all these asymmetries has a $\chi^2/dof$
of $11.8/5$, corresponding to a probability of only $3.7\%$ for the asymmetry
data to be consistent with the SM hypothesis.
Such a large discrepancy mainly stems from the two most precise determinations
of $\sin^2 \theta_{eff}$, namely the measurement of $A_{LR}$ by SLD and
the measurement of the bottom forward-backward asymmetry $A_{FB}^{b}$ at LEP,
which give values on opposite sides of the average and differ by $3.2$ standard
deviation.  It is interesting to note that if such
a discrepancy is attributed to experimental origin and thus the
hadronic asymmetry measurements are not included in the global fit,
then a rather light Higgs boson around 50 GeV is indicated from the fit \cite{Chanowitz,electro-data},
which is in sharp contrast with the LEP II direct search limit of 114 GeV \cite{Barate}
and results in a compatible probability as low as $3\%$.
If we resort to new physics to solve this discrepancy, the new physics
effects must significantly modify the $Z b \bar{b}$ coupling while
maintain the $Z$-boson couplings to other fermions basically unchanged.
In this work we focus on the $Z b \bar{b}$ coupling and scrutinize the
supersymmetric effects.

In our analysis we choose to parameterize the $Z f \bar{f} $ interaction at $Z$-pole
in term of the parameter $\rho_f$ and effective electroweak mixing angle $\sin^2
\theta^f_{eff} $ \cite{Veltman,Jegerlehner}:
\begin{eqnarray}
\Gamma_{Z f \bar{f} }^\mu &=&  (\sqrt{2} G_\mu \rho_f )^\frac{1}{2}
m_Z \gamma^\mu \left[  - 2 Q_f \sin^2 \theta_{eff}^f + I_3^f ( 1 -
\gamma_5 ) \right] \label{redefined}
\end{eqnarray}
This parametrization is preferred from the experimental point of
view because all the measured asymmetries are only dependent on
$\sin^2 \theta_{eff}^f$ and their precise measurements can
directly determine the value of $\sin^2 \theta_{eff}^f$. From the
combined LEP and SLD data analysis, the fitted values of $\rho_f$
and $\sin^2 \theta_{eff}^f$ agree well with their SM predictions for
leptons and light quarks, but for the bottom quark their fitted
values are respectively $1.059 \pm 0.021$ and $0.281 \pm 0.016$
(with correlation coefficient 0.99), which significantly deviate
from their SM predictions of 0.994 and 0.233 (for $m_t =174$ GeV
and  $m_h = 115$ GeV) and leads to $\chi^2/dof = 9.62/2$
(corresponding to a compatible probability of $0.8\%$). To best
fit the experimental data, $\rho_b$ and $\sin^2\theta^b_{eff}$
should be enhanced by about $6.5\%$ and $20\%$, respectively.
While we can envisage that the supersymmetric effects are not
usually so large, we want to figure out to what extent
supersymmetry can improve the situation. For this purpose, we
choose two popular supersymmetric models: the minimal
supersymmetric model (MSSM) \cite{Haber} and the next-to-minimal
supersymmetric model (NMSSM) \cite{Franke}.

For the NMSSM effects on $Z b \bar{b}$ coupling,
which have not been studied in the literature,
we will perform the calculation to one-loop level.
For the MSSM effects, which have been studied by many
authors \cite{Djouadi,Boulware,Cao}, we will renew the study
in the parametrization of $\rho_b$ and $\sin^2 \theta_{eff}^b$
(the previous studies usually examined the effects on the $Z$-width,
the ratio $R_b$ and the asymmetry $A_{FB}^b$).
For both the MSSM and NMSSM, we will
consider various current experimental constraints on the
parameter space, which are from the precision electroweak measurements,
the direct search for sparticles and Higgs bosons, the stability of the Higgs potential,
the cosmic dark matter relic density, and the muon g-2 measurement.

This paper is organized as follows. In Sec.II we introduce the general formula
for the calculation of $\rho_f$ and $\sin^2 \theta_{eff}^f $ and apply them to the MSSM and
NMSSM. In Sec.III we summarize the constraints considered in this work and
briefly discuss their characters. In Sec. IV and Sec. V
we perform numerical study for the corrections to $\rho_b$ and $\sin^2 \theta_{eff}^b $
in the MSSM and NMSSM, respectively. We will first show the characters of different
type corrections, then we will scan the whole SUSY parameter space to investigate
the compatibility of the supersymmetric predictions of $\rho_b$ and $\sin^2 \theta_{eff}^b $
with their experimental results. Finally, in Sec. VI we conclude our work
with an outlook on the possibility of solving the $Z b \bar{b}$ anomaly.

\section{general formula to calculate $\rho_f$ and $\sin^2
\theta_{eff}^f$}

In the SM with the input parameters the Fermi constant $G_F$, the
fine-structure constant $\alpha$, $Z$-boson mass $m_Z$ and fermion
masses $m_f$, the electroweak mixing angle $ s_W = \sin \theta_W $
is determined at loop level by  \cite{Sirlin,Hollik,Denner}
\begin{eqnarray}
s_W^2= \frac{1}{2} \left ( 1 - \sqrt{ 1 - \frac{ 4 \pi
\alpha}{\sqrt{2} G_\mu m_Z^2} \frac{1}{ 1 - \Delta r  } }  \ \right
) \label{sw2}
\end{eqnarray}
where $\Delta r $ is given by
\begin{eqnarray}
\Delta r = \frac{\hat{\Sigma}^W (0)}{m_W^2} + \frac{\alpha}{4 \pi
s_W^2} \left( 6 + \frac{7 - 4 s_W^2}{2 s_W^2} \ln ( 1 - s_W^2)
\right) + 2 \delta^v + \delta^b \label{deltar}
\end{eqnarray}
with $\hat{\Sigma}^W $ denoting the renormalized $W$-boson
self-energy, $ \delta^v $ and $\delta^b $ being the vertex
correction and box diagram correction to $\mu$ decay $\mu \to
\nu_\mu e \bar{\nu}_e$, respectively. To get a more precise
numerical result for $s_W^2$, one can iterate Eqs.(\ref{sw2}) and
(\ref{deltar}) a few times.

With the $s_W$ defined above, the effective $Z f \bar{f}$ coupling at
Z-pole takes the following form \cite{Jegerlehner,Hollik}
\begin{eqnarray}
\Gamma_{Z f \bar{f} }^\mu &=& \left(\sqrt{2} G_\mu ( 1 - \Delta r)\right)^{\frac{1}{2}}
m_Z \gamma^\mu \left \{ v_f - a_f \gamma_5 + \delta
v_f - \delta a_f \gamma_5 \right. \nonumber \\
&& \left. - \frac{1}{2} \left[ \Sigma_Z^\prime (m_Z^2) + \delta
Z_2^Z \right] ( v_f - a_f \gamma_5 ) - 2 Q_f s_W^2 \Delta \kappa
\right \}, \label{original}
\end{eqnarray}
where $v_f  =  I_3^f - 2 Q_f s_W^2 $ and $a_f = I_3^f $ are respectively
the vector and axial vector coupling coefficients of $Z f \bar{f}$ interaction
at tree level, and $\delta v_f$ and $\delta a_f$ are their corresponding
corrections. $\Sigma_Z^\prime $ is the derivative of the
unrenormalized $Z$-boson self-energy $\Sigma_Z$ with respect to
the squared momentum $p^2$, and $\delta Z_2^Z$ is
the field renormalization constant of $Z$-boson given by
\begin{eqnarray}
\delta Z_2^Z = - \Sigma^\prime_\gamma (0) - 2 \frac{c_W^2 - s_W^2 }{
s_W c_W} \frac{\Sigma_{\gamma Z}(0)}{m_Z^2} + \frac{c_W^2 -
s_W^2}{s_W^2} \left ( \frac{Re \Sigma_Z (m_Z^2)}{m_Z^2} - \frac{Re
\Sigma_W (m_W^2)}{m_W^2} \right ),
\end{eqnarray}
and $\Delta \kappa $ is given by
\begin{eqnarray}
\Delta \kappa &=& \frac{c_W^2}{s_W^2} \left \{ \frac{\Sigma_Z
(m_Z^2) }{m_Z^2} -  \frac{\Sigma_W (m_W^2) }{m_W^2} -
\frac{s_W}{c_W} \frac{\Sigma_{\gamma Z} (m_Z^2) + \Sigma_{\gamma Z}
(0) }{m_Z^2} \right \}.
\end{eqnarray}
In Eq.(\ref{original}) the factor $\frac{1}{2} ( \Sigma_Z^\prime
(m_Z^2) + \delta Z_2^Z ) $  comes from the fact that the residue of
the renormalized Z propagator is different from 1, while the last
term enters due to $Z-\gamma$ mixing at $Z$-pole.

If we re-express $\Gamma_{Z f \bar{f} }^\mu $ in Eq.(\ref{original})
in term of $\rho_f$ and $ \sin \theta_{eff}^f$ as in
Eq.(\ref{redefined}), we get
\begin{eqnarray}
\rho_f &=& 1 + \delta \rho_{se} + \delta \rho_{f, v}, \\
\sin^2 \theta_{eff}^f & =& ( 1 + \delta \kappa_{se} + \delta
\kappa_{f,v} ) s_W^2,
\end{eqnarray}
with $\delta \kappa_{se}= \Delta \kappa $ and
\begin{eqnarray}
\delta \rho_{se} & =&
   \frac{\Sigma_Z (0) }{m_Z^2} - \frac{\Sigma_W (0) }{m_W^2}
    - 2 \frac{s_W}{c_W} \frac{\Sigma_{\gamma Z} (0)}{m_Z^2}
    + \frac{\Sigma_Z (m_Z^2) - \Sigma_Z (0)}{m_Z^2} -
    \Sigma^\prime_Z(m_Z^2); \nonumber \\
\delta \rho_{f, v}& =& 2 \frac{\delta a_f}{a_f} - 2 \delta^v - \delta^b;  \nonumber \\
\delta \kappa_{f, v} &=& \frac{a_f \delta v_f - v_f \delta a_f}{- 2
Q_f a_f s_W^2}. \label{definition}
\end{eqnarray}
In above equations the subscript `$se$' means the contribution from
the gauge boson self-energy which is flavor independent, and `$f,v$'
denotes the contribution from the vertex correction to $ Z f \bar{f}$
interaction. In practice, it is convenient to express $\delta
\rho_{f,v}$ and $\delta \kappa_{f,v}$ in term of $\delta g_L^f $ and
$\delta g_R^f$ respectively
\begin{eqnarray}
\delta \rho_{f, v}& =&  \frac{\delta g_L^f - \delta g_R^f }{a_f} - 2
\delta^v - \delta^b;  \quad \quad  \delta \kappa_{f, v} = \frac{ (
a_f - v_f ) \delta g_L^f + ( a_f + v_f ) \delta g_R^f}{- 4 Q_f a_f
s_W^2 } \label{drb}
\end{eqnarray}
where $ \delta g_{L,R}^f = \delta v_f \pm \delta a_f $ are the
corrections to $Z  f_L \bar{f}_L $ and $Z  f_R \bar{f}_R $
interactions, respectively. From above equations one can learn that
the correction to $\delta \rho_{f,v}$ is decided by the competition
of $\delta g_L^f$ and $\delta g_R^f$, while $\delta \kappa_{f,v}$ is
mainly determined by $\delta g_R^f$ due to $ (a_f + v_f)/(a_f - v_f)
\simeq 5.4$.

Noting that the Feynman rules for $Z$-boson couplings in SUSY models
usually differ from their corresponding rules in the SM by a minus
sign \cite{Haber,Franke}, $\Sigma_{\gamma Z}$ and $\delta
\kappa_{f,v} $ in the above formula should change sign if one uses
the Feynman rules in SUSY models. The self-energies and the vertex
corrections in SUSY models then include both the SM-particle loop
contributions and SUSY-particle loop contributions. Since the
SM-particle contributions are well known, in Appendix A and B we
only list the one-loop expressions for the SUSY contributions. The
only subtlety one should note is to avoid the double-counting of the
Higgs contributions. This problem arises due to the following
reason. On the one hand, the SM values of $\rho_{b}$ and $\sin^2
\theta_{eff}^b $ are known to higher orders, and one usually
incorporates such high-order SM effects when performing numerical
calculations in SUSY models. On the other hand, because the SUSY
Higgs sector is quite different from the SM, one cannot get the SUSY
Higgs contributions simply by adding some additional terms to the SM
Higgs contributions. In our calculation in SUSY models, to avoid the
double-counting of the Higgs contributions, we first subtract the SM
Higgs contributions from their SM values (calculated by the codes
TOPAZ0 \cite{Montagna} and ZFITTER \cite{Bardin}), and then we add
the full one-loop contributions from the SUSY Higgs bosons and
sparticles.

\section{constraints on SUSY parameters}

Before we proceed to discuss the SUSY corrections to $Zb\bar b$
coupling in the MSSM and NMSSM, we take a look at the SUSY
parameters involved in our calculations. From the expressions of $Z
f \bar{f}$ vertex correction listed in Appendix B, one can learn
that the SUSY- EW correction depends on the masses and the mixings
of top squarks, bottom squarks, charginos and neutralinos, the
SUSY-QCD vertex correction depends on gluino mass and the masses and
the chiral mixing of bottom squarks, and the Higgs-mediated vertex
correction depends on the masses and the mixings of Higgs bosons.
The expressions of the gauge boson self-energies listed in Appendix
A indicate that the SUSY correction also depends on the masses of
sleptons and the first-two generation squarks. About these SUSY
parameters, we consider the following constraints
\begin{itemize}
\item[(1)] Constraints from the direct search for the sparticles at LEP and Tevatron  \cite{Yao}
\begin{eqnarray*}
&& m_{\tilde{\chi}_1^0} > 41 {\rm ~GeV}, \quad m_{\tilde{\chi}_2^0} > 62.4  {\rm ~GeV},
   \quad m_{\tilde{\chi}_3^0} > 99.9  {\rm ~GeV},  \quad  m_{\tilde{\chi}^\pm} > 94  {\rm ~GeV}, \\
&& m_{\tilde{e}} > 73  {\rm ~GeV},  \quad m_{\tilde{\mu}} > 94  {\rm ~GeV},
   \quad m_{\tilde{\tau}} > 81.9 {\rm  GeV}, \quad m_{\tilde{q}} > 250  {\rm ~GeV}, \\
&& m_{\tilde{t}} > 89  {\rm ~GeV}, \quad m_{\tilde{b}} > 95.7  {\rm ~GeV},
  \quad m_{\tilde{g}} > 195  {\rm ~GeV},
\end{eqnarray*}
where $m_{\tilde{\chi}^0_i}$ denote the masses of the neutralinos
and $m_{\tilde{q}}$ denotes the masses for the first two generation
squarks.

\item[(2)] Constraint from the direct search for Higgs boson at LEP  \cite{Higgs}.
This constraint can limit the values of $m_A$, $\tan \beta$ and the
masses and the chiral mixing of top squarks. In case of large
$\tan \beta$, it can also put constraints on the masses and the
mixing of bottom squarks. Generally speaking, this constraint
requires the product of two top squark masses, $m_{\tilde{t}_1}
m_{\tilde{t}_2}$,  should be much larger than
$m_t^2$  \cite{Higgs-theory}.

\item[(3)] Constraint from the theoretical requirements that there is no
Landau pole for the running Yukawa couplings $Y_b$ and $Y_t$ below the
GUT scale, and that the physical minimum of the Higgs potential with
non-vanishing $ \langle H_u \rangle$ and $ \langle H_d \rangle $ is
lower than the local minima with vanishing $ \langle H_u \rangle$
and $ \langle H_d \rangle $.

\item[(4)] Constraints from precision electroweak observalbes such as
$\rho_{lept}$, $\sin^2 \theta_{eff}^{lept}$, $\rho_c$, $\sin^2
\theta_{eff}^c$ and $M_W$.  These constraints are equivalent to
those from the well known $\epsilon_i (i=1,2,3) $
parameters \cite{Altarelli} or $S$, $T$ and $U$
parameters \cite{Peskin}. The measured values of these observables
are \cite{2005ema}
\begin{eqnarray}
&&\rho_{lept} = 1.0050 \pm 0.0010, \quad \sin^2
\theta_{eff}^{lept} = 0.23153 \pm 0.00016,  \nonumber \\
&&\rho_c = 1.013 \pm 0.021, \quad  \sin^2 \theta_{eff}^{c} =
0.2355 \pm 0.0059, \quad  M_W = 80.403 \pm 0.029  {\rm ~GeV}, \nonumber
\end{eqnarray}
and their SM fitted values are $\rho_{lept}^{SM} = 1.0051 $, $
\sin^2 \theta_{eff}^{lept, SM} = 0.23149 $,  $ \rho_c^{SM}  = 1.0058
$,  $ \sin^2 \theta_{eff}^{c} = 0.2314 $ and $M_W = 80.36$ GeV for
$m_t = 173$ GeV and $m_h = 111$ GeV. In our  calculations we require
the theoretical predictions to agree with the experimental
values at $2\sigma$ level.

\item[(5)] Constraint from $R_b = \Gamma (Z \to b \bar{b} ) / \Gamma
( Z \to hadrons ) $.  The measured value of $R_b$ is $R_b^{exp} =
0.21629 \pm 0.00066 $ and its SM prediction is $R_b^{SM} = 0.21578 $
for $m_t = 173$ GeV \cite{Yao}. In our analysis, we require
$R_b^{SUSY}$ is within the $2 \sigma$ range of its experimental value.

\item[(6)] Constraint from the relic density of cosmic dark matter,
i.e. $ 0.0945 < \Omega h^2 < 0.1287 $ \cite{dmconstr}. This
constraint can rule out a broad parameter region for guagino masses
$M_{1,2}$, $\mu$ parameter, $m_A$ and $\tan \beta$ \cite{darkmatter}.

\item[(7)] Constraint from the muon anomalous magnetic momentum, $a_\mu$.
Now both the theoretical prediction and the experimental measurement of
$a_\mu$ have reached a remarkable precision, which show a significant
deviation
$a_\mu^{exp} - a_\mu^{SM} = ( 29.5 \pm 8.8 ) \times 10^{-10} $ \cite{Miller}.
In our analysis we require the SUSY effects to account for such difference
at $2 \sigma$ level.

\end{itemize}

Note that in our analysis we do not include the constraints from $B$ physics,
like $b \to s \gamma$ \cite{bsr} and $B_s-\bar{B_s}$ mixing \cite{Ball},
because these constraints are sensitive to squark flavor mixings
which are irrelevant to our discussion.

Among the  constraints listed above, the
constraints (4) and (5), especially the observables  $M_W$,
$\rho_{lept}$, $\sin^2 \theta_{eff}^{lept}$ and $R_b$, are most
relevant to our study of $\rho_b$ and $\sin^2 \theta_{eff}^b$. Let
us look at these constraints in more details.

First, the precise measurements of $M_W$, $\rho_{lept}$ and $\sin^2
\theta_{eff}^{lept}$ stringently constrain $\delta \rho_{se}$,
$\delta \kappa_{se}$ and the gaugino loop contributions to $\delta
\rho_{b,v}$ and $\delta \kappa_{b,v}$. The approximate forms of the
SUSY corrections to $M_W$, $\delta \rho_{se} $ and $\delta
\kappa_{se}$ \cite{Heinemeyer} in case of heavy sparticles are given
by
\begin{eqnarray}
\frac{\delta M_W}{M_W} &= & \frac{s_W^2}{c_W^2 - s_W^2}
\frac{\delta (\Delta r)}{ 2 ( 1 - \Delta r ) } \simeq -
\frac{c_W^2}{c_W^2 - s_W^2} \frac{\Delta \rho}{2},  \nonumber \\
\delta \rho_{se} & \simeq & \Delta \rho,  \nonumber \\
\delta \kappa_{se} & \simeq & \frac{c_W^2}{s_W^2} \Delta \rho,
\end{eqnarray}
where
\begin{eqnarray}
\Delta \rho & =& \frac{\Sigma_Z (0) }{m_Z^2} - \frac{\Sigma_W (0)
}{m_W^2} - 2 \frac{\sin \theta_W}{\cos \theta_W}
\frac{\Sigma_{\gamma Z} (0) }{m_Z^2}
\end{eqnarray}
is the correction to the classical $\rho $ parameter \cite{Veltman}
and is only sensitive to the mass spectrum of the third generation
squarks. Through the above relations the precisely measured $M_W$
then stringently restricts $\Delta \rho$ (of order $O(10^{-4})$) and
subsequently restricts $\delta \rho_{se} $ and $\delta \kappa_{se}$.
This restriction together with the precisely determined
$\rho_{lept}$ and $\sin^2 \theta_{eff}^{lept}$ stringently
constrains the magnitude of $\delta \rho_{l,v}$ and $\delta
\kappa_{l,v}$ defined in Eq.(\ref{definition}) to be below
$O(10^{-4})$. Since the gaugino loop effects in $\delta \rho_{b,v} $
and $\delta \kappa_{b,v}$ are strongly correlated with $\delta
\rho_{l,v}$ and $\delta \kappa_{l,v}$ (the main difference is caused
by the mass difference between sleptons and squarks),  the gaugino
loop contributions to $\delta \rho_{b,v}$ and $\delta \kappa_{b,v}$
are also suppressed, which are found to be below $5 \times 10^{-4}$
from our numerical calculations.

For the constraint from the precision observable $R_b$, an interesting character
is that it does not stringently constrain the magnitude of $\delta v_b $ and $\delta a_b$,
but it favors the relation $\delta v_b \sim -1.44 \delta a_b$, which can
be seen from the expression of the radiative correction to
$R_b$ \cite{Djouadi,Boulware,Cao}
\begin{eqnarray}
\delta R_b& \simeq & \frac{2 R_b^{SM}(1-R_b^{SM})
}{v_b^2(3-\beta^2)+2a_b^2\beta^2} \big[v_b(3-\beta^2) \delta v_b
 + 2a_b\beta^2 \delta a_b \big] \propto ( \delta v_b + 1.44  \delta a_b )
\end{eqnarray}
with $\beta = \sqrt{ 1 - m_b^2/m_Z^2} $ being the velocity of bottom
quark in $Z$ decay.

Now we turn to the constraint from the muon anomalous magnetic
momentum. To get an intuitive understanding of this constraint, we
look at a simple case of the MSSM that all the gaugino masses and
soft-breaking masses in smuon sector have a common scale $M$. In
this case, $a_\mu^{SUSY} $ is approximated by \cite{Ibrahim}
\begin{eqnarray}
a_\mu^{SUSY} \simeq 13 \times 10^{-10} \left ( \frac{100  {\rm ~GeV}}{M}
\right )^2 \tan \beta\  sign(\mu).
\end{eqnarray}
The gap between $a_\mu^{SM}$ and $a_\mu^{exp}$  then prefers a
positive $\mu$, and constrains the product  $ \left ( \frac{100  {\rm ~GeV}}{M}
\right )^2 \tan \beta$ in the range [1.0,3.6] at $2\sigma$ level.
So the SUSY scale can be higher for larger $\tan \beta$.

In our calculations we use the code NMSSMTools \cite{Ellwanger} to
generate the masses and the mixings for all sparticles and Higgs
bosons in the framework of the NMSSM with all known radiative
corrections included. There are two advantages in using this code.
One is that all the masses and the mixings in the MSSM can be easily
recovered if we set the parameters $\lambda = \kappa \simeq 0 $ and
$A_{\kappa}$ to be negatively small. The other is that it
incorporates the code MicrOMEGAs \cite{Belanger} which calculates
the relic density of cosmic dark matter. It should be noted that the
current version of NMSSMTools only includes the constraints (1),
(2), (3) and (6), and we extend it by including the constraints (4),
(5) and (7). We note that the muon anomalous magnetic momentum was
recently calculated in the NMSSM \cite{Domingo} and our calculations
agree with theirs.

\section{ One-loop corrections to $\rho_b$ and $\sin^2 \theta_{eff}^b$ in MSSM }

In this section we investigate $\rho_b$ and $\sin^2
\theta_{eff}^b$ to one-loop level in the MSSM.
As discussed above, the self-energy corrections to these two observables
are generally small and thus we mainly scrutinize
the vertex corrections which include the SUSY-EW corrections,
the SUSY-QCD corrections and the Higgs-mediated vertex corrections.
We pay special attention to the cases where the magnitudes of the
corrections are large, and show that $\tan \beta$ is crucial in
enhancing the vertex corrections. Our analysis is organized as follows:
we first investigate the characters of the vertex corrections to get an
intuitive understanding of them, then by scanning over the MSSM parameter
space, we study the compatibility of the MSSM predictions for $\rho_b$ and
$\sin^2 \theta_{eff}^b$ with their experimental results.

The SM input parameters involved in our calculations are taken from \cite{Yao},
which are $\alpha = 1./128.93$, $G_F = 1.16637 \times 10^{-5}$,
$\alpha_s (m_Z) = 0.1172$, $m_Z = 91.1876$ GeV, $m_b (m_b )= 4.2$ GeV
and $m_t = 172.5$ GeV.

\subsection{Characters of vertex corrections in MSSM}

As for the SUSY-EW contribution to $\delta \rho_{b,v}$ and $\delta
\kappa_{b,v}$, the parameters involved are guagino masses $M_{1,2}$,
Higgsino mass $\mu$, $\tan \beta =v_2/v_1$ with $v_{1,2}$
being the vacuum expectation values of the Higgs fields, the
soft-breaking masses $M_{Q_3}$, $M_{U_3}$, $M_{D_3}$, and the
coefficients of the trilinear terms $A_t$ and $A_b$. The first four
parameters enter the mass matrices of neutralinos and charginos, and
the last seven parameters affect the masses and the chiral mixings
of the third generation squarks \cite{Haber}.

As discussed in the preceding section, the gaugino loop contribution
is small, and hence we only discuss the Higgsino loop contribution.
The magnitude of such Higgsino loop contribution is sensitive to
$\tan \beta $, the Higgsino mass $\mu$, and the masses and the chiral
mixings of the third generation squarks. There are two characters
for this contribution. One is that, due to the fact that the bottom
Yukawa coupling $Y_b$ is proportional to $ 1/\cos \beta $, the
contribution can be potentially large in case of large $\tan \beta$
and small $\mu$.  The other is that the contribution is moderately
sensitive to the chiral mixings of the third generation squarks,
and potentially large contribution comes from the case where the
mixing is small and the component of the lighter squark is
dominated by the left-handed squark \cite{Boulware}.
To illustrate these characters we consider three cases in the
squark sector:
\begin{itemize}
\item[(I)] $M_S= M_{Q_3} = M_{U_3} = M_{D_3} = 400$ GeV,
           $A_t = A_b = 800$ GeV;
\item[(II)] $M_{Q_3} = 200$ GeV,
            $M_{U_3} = M_{D_3} = 600$ GeV,
            $A_t = A_b = 800$ GeV;
\item[(III)] $M_{Q_3} = 600$ GeV,
             $M_{U_3} = M_{D_3} = 200$ GeV,
             $A_t = A_b = 800$ GeV,
\end{itemize}
and fix other SUSY parameters as
\begin{eqnarray}
M_1 = 75  {\rm ~GeV}, \quad  M_2 = 150  {\rm ~GeV}, \quad m_A = 500  {\rm ~GeV}, \quad
M_{SUSY} = 1  {\rm ~TeV}, \label{parameters}
\end{eqnarray}
where $M_{SUSY}$ denotes the soft-breaking parameters for sleptons
and the first-two generation squarks. Case-I corresponds to maximal
chiral mixing case, Case-II is the small mixing case with the
component of the lighter squark dominated by the left-handed squark
and Case-III is also the small mixing case but with the component of
the lighter squark dominated by the right-handed squark.

%%%%%%%%%%%%%%%%%%%%%%%%%%%%%%%%%%%%%%%%%%%%%%%%%%%%%%%%%%%%%%%%%%%%
\begin{figure}[tbp]
\epsfig{file=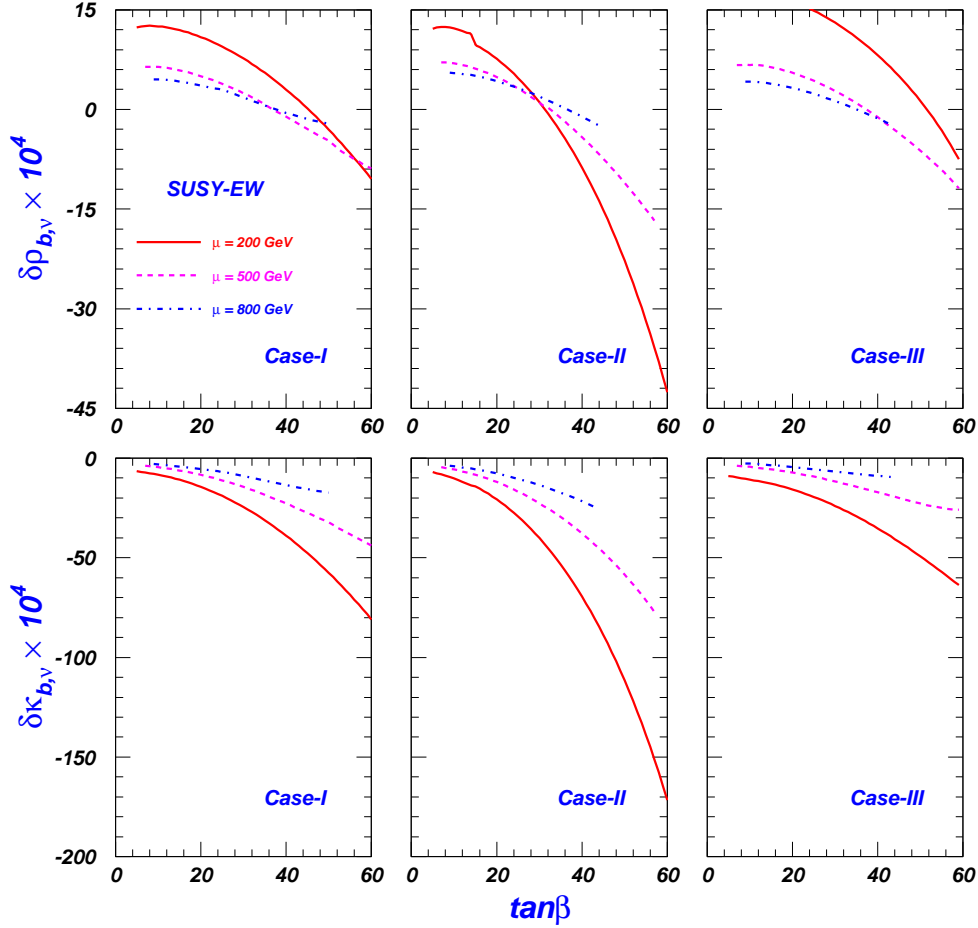,width=13cm}
\vspace{-0.5cm}
\caption{\small SUSY-EW contributions to $\delta \rho_{b,v}$ and
          $\delta \kappa_{b,v}$ with constraints (1-5).}
\label{SUSY-EW1}
\end{figure}
%%%%%%%%%%%%%%%%%%%%%%%%%%%%%%%%%%%%%%%%%%%%%%%%%%%%%%%%%%%%%%%%%%%%

In Fig.\ref{SUSY-EW1} we show the dependence of the SUSY-EW
contribution to $\delta \rho_{b,v} $ and  $\delta \kappa_{b, v}$ on
$\tan \beta $ in the three cases. One can see that both $\delta
\rho_{b,v}$ and $\delta \kappa_{b,v}$ are sensitive to $\tan \beta$.
As $\tan \beta$ increases, $\delta \rho_{b,v}$ and $\delta
\kappa_{b,v}$ get more negative contributions and, for small $\mu$,
they become negative with sizable magnitudes. This behavior can be
understood as following. As $\tan \beta$ gets large, the bottom
Yukawa coupling increases and the correction to the right-handed $Z
b \bar{b}$ coupling $\delta g_R^b$ increases positively, and then
$\delta \rho_{b,v}$ and $\delta \kappa_{b,v}$ get more negative
contribution from the increasing $\delta g_R^b$(see Eq.(\ref{drb})
and also $\delta g_R^b$ in Appendix B). One also see from these
figures that the magnitude of $\delta \kappa_{b,v}$ is usually
larger than $\delta \rho_{b,v}$. The factor $\sin^2 \theta_W $ in
the denominator of $\delta \kappa_{b,v} $ (see Eq.(\ref{definition})
) can to a large extent account for this.

Note that in these figures we only plot our results within the range
of $\tan \beta $ that survives the constraints (1-5).
The constraint (7), i.e. the muon anomalous magnetic moment, can in
principle also limit $\tan \beta $. But this constraint relies on
the mass scale of smuon, $M_{SUSY}$ in Eq.(\ref{parameters}), which
$\rho_b$ and $\sin^2 \theta_{eff}^b $ are not sensitive to, so we do
not apply it in plotting these figures. Our numerical results
indicate that the muon anomalous magnetic moment allows for a vast
region of $M_{SUSY}$ and $\mu$ where $\tan \beta $ can be as large
as 60,  and hence the sizable SUSY-EW corrections to $\rho_b$ and
$\sin^2 \theta_{eff}^b$ are possible. For example, with the
parameters in Eq.(\ref{parameters}), the range of $\tan \beta $
allowed by the muon $g-2$ is $\tan \beta \geq 25 $ for $\mu = 200$
GeV, $\tan \beta \geq 33$ for $\mu = 500$ GeV,  and $\tan \beta \geq
44 $ for $\mu = 800$ GeV. If we choose $M_{SUSY} = 0.5$ TeV, these
allowed ranges are correspondingly given by $ 7 \leq \tan \beta \leq
57 $, $12 \ \leq \tan \beta \leq 71 $ and $\tan \beta \geq 14$.

\begin{figure}[tbp]
\epsfig{file=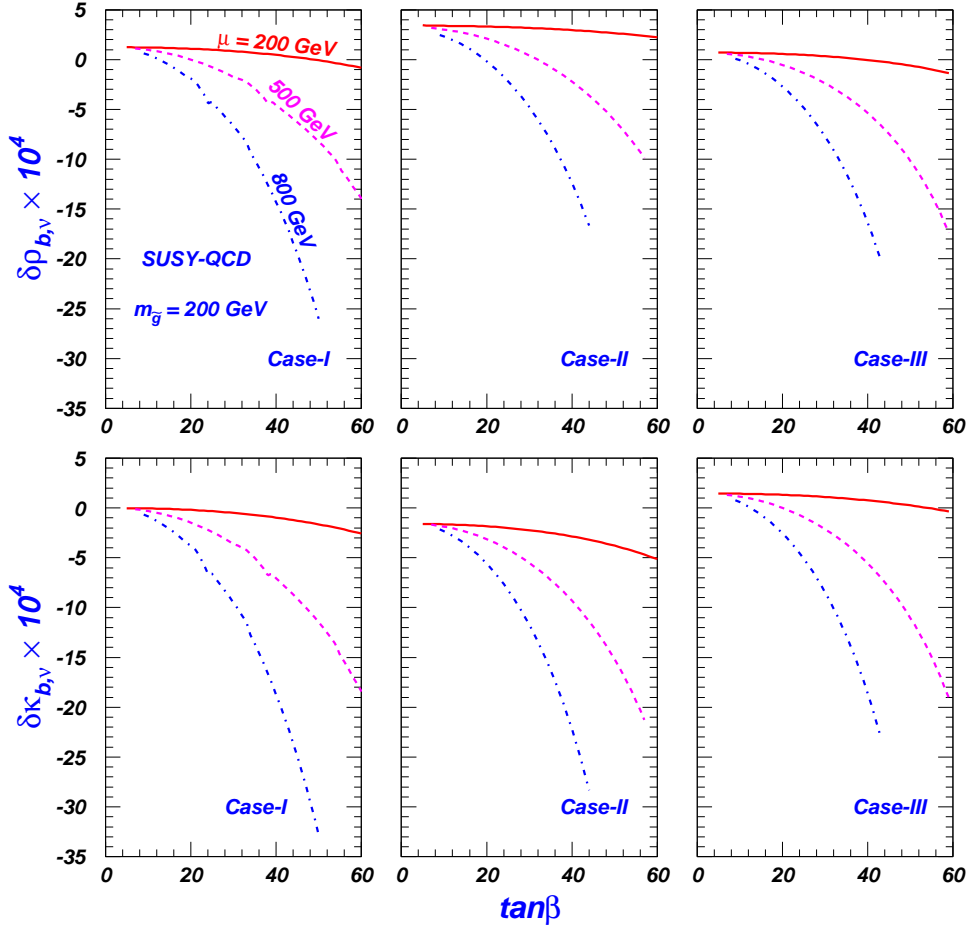,width=13cm}
\vspace{-0.5cm}
\caption{\small SUSY-QCD contributions to $\delta \rho_{b,v}$ and
         $\delta \kappa_{b,v}$ with constraints (1-5).}
\label{SUSY-QCD1}
\end{figure}

Next we discuss the SUSY-QCD corrections. The relevant parameters
are gluino mass and $M_{Q_3}$, $M_{D_3}$ and $X_b = ( A_b - \mu \tan
\beta ) $ which enter the mass matrix of the bottom squarks. From
the large strength of the strong coupling, $g_s (m_Z ) \simeq 1.2
\simeq 50 \times Y_b^{SM} $, one may naively postulates that the
SUSY-QCD contributions to $\delta \rho_{b,v}$ and $\delta
\kappa_{b,v} $ should be much larger than the Higgsino loop
contributions in case of $m_{\tilde{g}} \simeq \mu $ and $\tan \beta
\ll 50$. However, our numerical results show that in case of small
sbottom chiral mixing the SUSY-QCD contributions to $\delta
\rho_{b,v}$ and $\delta \kappa_{b,v}$ are negligibly small. The
underlying reason is that for the SUSY-QCD corrections there is a
strong cancellation between different diagrams in case of small
sbottom chiral mixing, which can be seen from the expressions of
$\delta g_{L,R}^b $ listed in Appendix B. It should be noted that
such a cancellation can be alleviated for a large sbottom mixing, or
equivalently, a large term $\mu \tan \beta $ appeared in the
non-diagonal elements of sbottom mass matrix (we checked this from
numerical calculations). So the contribution may be sizable in case
of large $\mu \tan \beta $, as shown in Fig.\ref{SUSY-QCD1}.

%%%%%%%%%%%%%%%%%%%%%%%%%%%%%%%%%%%%%%%%%%%%%%%%%%%%%%%%%%%%%%%%%%%%
%%%%%%%%%%%%%%%%%%%%%%%%%%%%%%%%%%%%%%%%%%%%%%%%%%%%%%%%%%%%%%%%%%%%

Compared with the Higgsino loop corrections, the SUSY-QCD
contributions in Fig.\ref{SUSY-QCD1} exhibit
a similar behavior with respect to $\tan \beta$.
The difference is that the most sizable effects come from Case-I
(maximal sbottom mixing case) with large $\mu$, instead of Case-II
with small $\mu$ for the Higgsino loop corrections.

\begin{figure}[tbp]
\epsfig{file=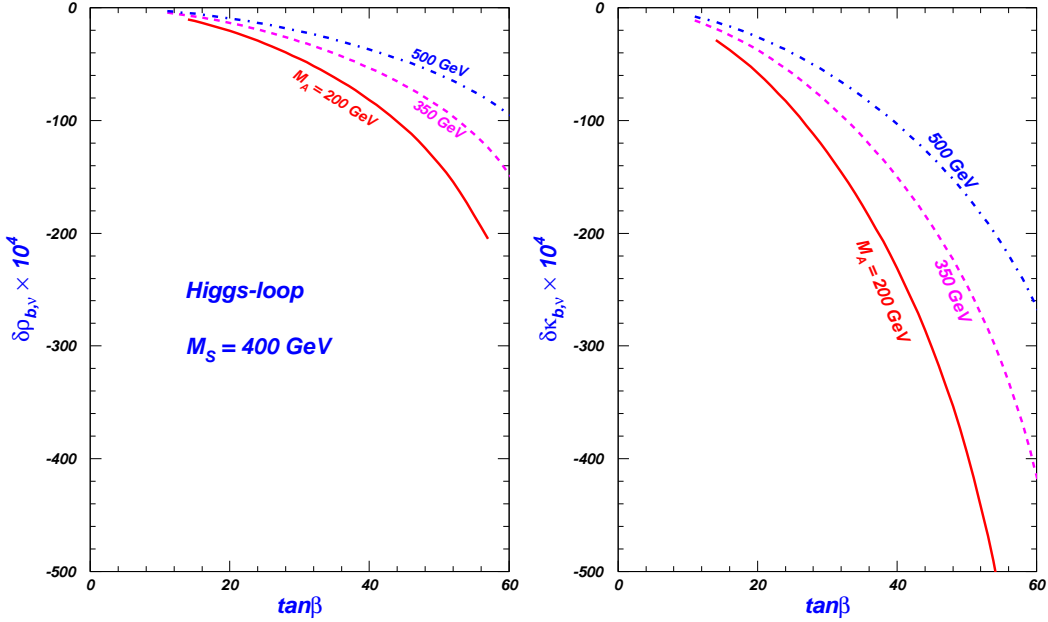,width=14cm}
\vspace{-0.5cm}
\caption{\small Higgs loop contributions to $\delta \rho_{b,v}$ and
         $\delta \kappa_{b,v} $ in Case-I with constraints (1-5).}
\label{SUSY-Higgs}
\end{figure}

Finally, we consider the Higgs loop contributions to $\delta
\rho_{b,v}$ and $\delta \kappa_{b,v}$ \cite{Logan}. To calculate
this part of contribution, we need to know the masses and the mixing
of the Higgs bosons, which are determined by $m_A$ and $\tan \beta $
at tree-level, and also by the soft-breaking masses for the third
generation squarks if the important loop correction to the Higgs
boson masses is taken into account. As shown in
Fig.\ref{SUSY-Higgs}, the contributions exhibit a similar dependence
on $\tan \beta$, and the significant contribution comes from the
case of small $m_A$ and large $\tan \beta$. We checked that the
results in Fig.\ref{SUSY-Higgs} are not sensitive to $\mu $ or
$M_S$, and also not sensitive to the choice of different case (
Case-I, Case-II or Case-III).

%%%%%%%%%%%%%%%%%%%%%%%%%%%%%%%%%%%%%%%%%%%%%%%%%%%%%%%%%%%%%%%%%%%%
%%%%%%%%%%%%%%%%%%%%%%%%%%%%%%%%%%%%%%%%%%%%%%%%%%%%%%%%%%%%%%%%%%%%

>From the above figures one can infer that among the three types of
corrections, the potentially largest correction comes from the
Higgs loops, which can reach $2\%$ for $\rho_b$ and $6\%$ for
$\sin^2 \theta_{eff}^b$. Such large corrections reach the current
experimental sensitivity since the current experimental
measurements are $\rho_b^{exp} = 1.059 \pm 0.021$ and $ \sin^2
\theta_{eff}^{b, exp} = 0.281 \pm 0.016 $.

%%%%%%%%%%%%%%%%%%%%%%%%%%%%%%%%%%%%%%%%%%%%%%%%%%%%%%%%%%%%%%%%%%%%
\begin{figure}[tbp]
\epsfig{file=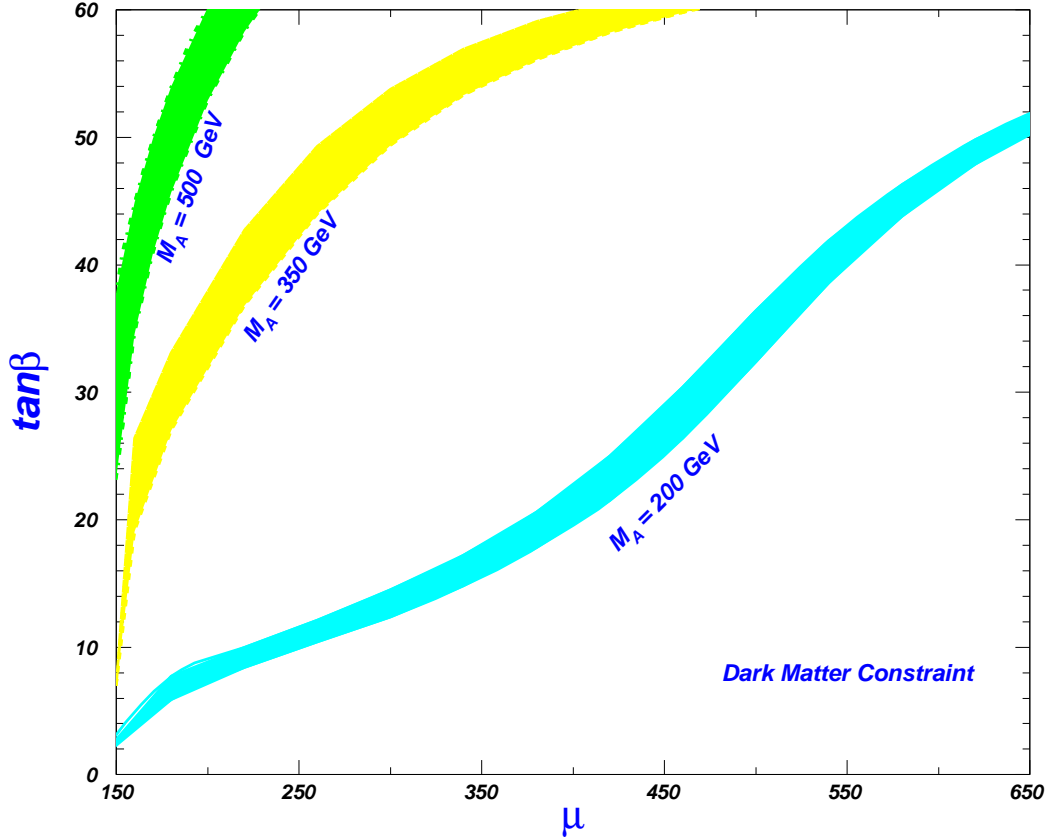,width=14cm}
\vspace{-0.5cm}
\caption{\small The shaded regions are allowed by the cosmic dark matter
          relic density at $2\sigma$ level.
          Other relevant SUSY parameters are fixed as in Case-I
          and in Eq.(\ref{parameters}).}
\label{dark matter}
\end{figure}
%%%%%%%%%%%%%%%%%%%%%%%%%%%%%%%%%%%%%%%%%%%%%%%%%%%%%%%%%%%%%%%%%%%%

Before we end this section, we would like to point out that in the
large $\tan \beta $ limit the relic density of cosmic dark matter
allows the possibility of small $\mu$ or small $m_A$ (but not both
small). This can be seen from Fig.\ref{dark matter}, where we show
the allowed regions in the plane of $\tan \beta $ versus $\mu$ for
different $m_A$. In plotting this figure, we choose Case-I and fix
other related parameters in Eq.(\ref{parameters}). Fig.\ref{dark
matter} implies that the SUSY-EW contribution and the Higgs-loop
contribution to $\delta \rho_{b,v}$ and $\delta \kappa_{b,v}$ cannot
simultaneously reach their maximal values.

%%%%%%%%%%%%%%%%%%%%%%%%%%%%%%%%%%%%%%%%%%%%%%%%%%%%%%%%%%%%%%%%%%%%
\begin{figure}[tbp]
\epsfig{file=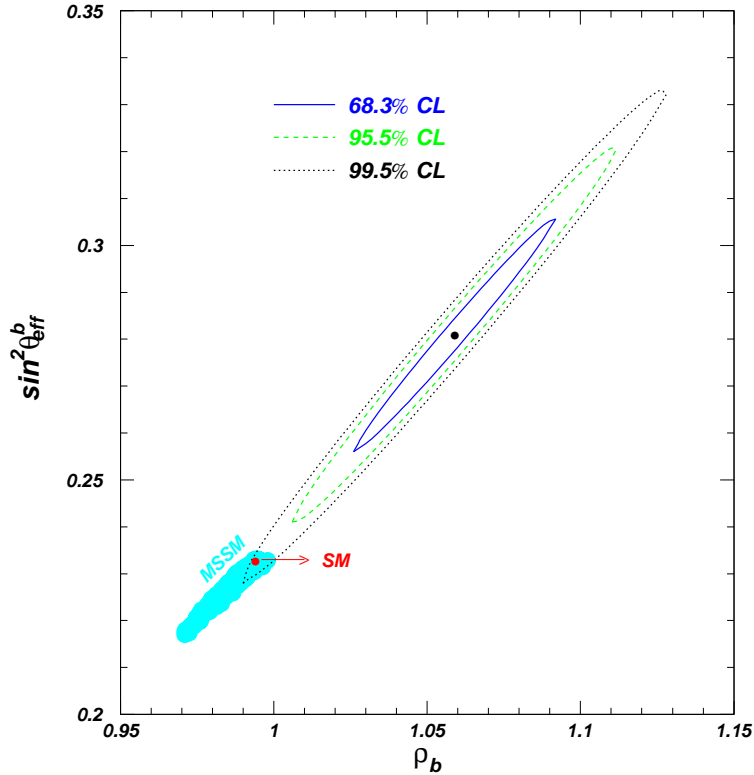,width=10cm} \vspace{-0.5cm} \caption{\small The
MSSM and SM predictions for  $\rho_b$
                and $\sin^2 \theta_{eff}^b$, compared with the
                LEP/SLD data at $68\%$, $95.5\%$ and $ 99.5\%$
                confidence level. The SM prediction
         $\rho_b^{SM} = 0.994 $ and $\sin^2 \theta_{eff}^{b, SM} = 0.233$
         is obtained with $m_t =174$ GeV  and  $m_h = 115$ GeV.
         The MSSM predictions are from a scan (a sample of one million)
         over the parameter space.}
\label{contours}
\end{figure}
%%%%%%%%%%%%%%%%%%%%%%%%%%%%%%%%%%%%%%%%%%%%%%%%%%%%%%%%%%%%%%%%%%%%

\subsection{MSSM predictions for  $\rho_b$ and $\sin^2 \theta_{eff}^b$}

As mentioned above, the extracted values of $\rho_b$ and $\sin^2
\theta_{eff}^b$ from combined LEP and SLD data analysis are
respectively $1.059 \pm 0.021$ and $0.281 \pm 0.016$ with
correlation coefficient 0.99 \cite{2005ema}. This result is shown
in Fig.\ref{contours} with the three ellipses corresponding to
$68\%$, $95.5\%$ and $ 99.5\%$ confidence level (CL),
respectively. Noting that the SM predictions are $\rho_b^{SM}=
0.994 $ and $\sin^2 \theta_{eff}^{b SM} = 0.233 $, one may infer
that large positive corrections to $\rho_b $ and $\sin^2
\theta_{eff}^b$ are needed to narrow the gap between the
experimental data and the SM prediction. As discussed in the
preceding section, the MSSM corrections can be sizable for large
$\tan\beta$, which, however, are negative and thus cannot narrow
the gap. To figure out to what extent the MSSM predictions can
agree with the experiment, we consider all the constraints
discussed in Sec. III and scan over the SUSY parameter space:
\begin{eqnarray}
&&0 < M_{1}, M_{2}, M_3, \mu, M_{Q_3}, M_{U_3}, M_{D_3}, M_A,
M_{SUSY} \leq 1 {\rm ~TeV}, \nonumber \\
&& -3 {\rm ~TeV} \leq A_t, A_b \leq 3 {\rm ~TeV}, \quad \quad 1 < \tan
\beta \leq 60, \label{region}
\end{eqnarray}
Based on a twenty billion sample, we find the best
MSSM predictions are $ \rho_b = 0.9960$ and $\sin^2 \theta_{eff}^b = 0.2328$,
which give a $\chi^2/dof = 9.07/2$ when compared with the experiment data.
If we do not consider the dark matter constraint,
the best MSSM predictions are $\rho_b = 0.99737$ and
$\sin^2 \theta_{eff}^b =0.2336 $, which give a $\chi^2/dof = 8.77/2$.
Moreover, we find that such a best case happens when $\mu, m_A, m_{\tilde{g}} \sim 1$
TeV so that the three types of vertex corrections are suppressed.

\section{ One-loop predictions for $\rho_b$ and $\sin^2 \theta_{eff}^b$
          in NMSSM }

\subsection{Introduction to the NMSSM}
As a popular extension of the MSSM, the NMSSM provides an elegant
solution to the $\mu$-problem via introducing a singlet Higgs
superfield $\hat{S}$, which naturally develops a vacuum expectation
value of the order of the SUSY breaking scale and gives rise to the
required $\mu$ term. Another virtue of the NMSSM is that it can
alleviate the little hierarchy problem since the theoretical upper
bound on the SM-like Higgs boson mass is pushed up and the LEP II
lower bound on the Higgs boson mass is relaxed due to the suppressed
$ZZh$ coupling or the suppressed decay $h \to b\bar b $
\cite{Dermisek}. Since the NMSSM is so well motivated, its
phenomenology has been intensively studied in recent years, such as
its effects in Higgs physics \cite{NMSSM-Higgs}, neutralino physics
\cite{NMSSM-Neutralino}, B-physics \cite{NMSSM-B} as well as squark
physics \cite{NMSSM-Sq}. In the following we recapitulate the basics
of the NMSSM with emphasis on its difference from the MSSM.

The superpotential of the NMSSM takes the form \cite{Franke,Ellwanger}
\begin{eqnarray}
W & = & \lambda \varepsilon_{ij} \hat{H}_u^i \hat{H}_d^j \hat{S}  +
\frac{1}{3} \kappa  \hat{S}^3 + h_u \varepsilon_{ij}
\hat{Q}^i \hat{U} \hat{H}_u^j - h_d \varepsilon_{ij}
\hat{Q}^i \hat{D} \hat{H}_d^j -h_e \varepsilon_{ij}
\hat{L}^i \hat{E} \hat{H}_d^j
\label{Superpotential}
\end{eqnarray}
where $\hat S$ is the singlet Higgs superfield, and
$\varepsilon_{12} = - \varepsilon_{21} = 1 $.
For the soft SUSY breaking terms, we take
\begin{eqnarray}
\label{vs} V_{\mbox{soft}} & = &
 \frac{1}{2} M_2 \lambda^a \lambda^a +\frac{1}{2} M_1 \lambda '\lambda '
 +m_d^2 |H_d|^2 + m_u^2 |H_u|^2+m_S^2 |S|^2  \nonumber \\
& & + m_Q^2 |\tilde{Q}|^2 + m_U^2 |\tilde{U}|^2 + m_D^2 | \tilde{D}|^2
  +m_L^2 |\tilde{L}|^2 + m_E^2 |\tilde{E}|^2 \nonumber \\
& &  + (\lambda A_\lambda \varepsilon_{ij} H_u^i H_d^j S + \mbox{h.c.})
     + (\frac{1}{3} \kappa A_\kappa S^3 + \mbox{h.c.}) \nonumber \\
& & + (h_u A_U \varepsilon_{ij} \tilde{Q}^i \tilde{U} H_u^j
   -h_d A_D \varepsilon_{ij} \tilde{Q}^i \tilde{D} H_d^j
   -h_e A_E \varepsilon_{ij} \tilde{L}^i \tilde{E} H_d^j +\mbox{h.c.})
\end{eqnarray}
With the above configuration of the model, the $\mu$ parameter is
given by $\mu = \lambda \langle S \rangle $ with $ \langle S \rangle
$ being the vacuum expectation value of $S$ field, and the $m_A$
parameter in the MSSM corresponds to the combination $m_A^2 =
\frac{2 \mu}{\sin 2 \beta } ( A_\lambda + \frac{\kappa \mu}{\lambda}
) $ (see Eq.(\ref{CP-odd})). So compared with the MSSM, the NMSSM
has three additional input parameters $\lambda$, $\kappa$ and
$A_\kappa$. These three parameters should be subject to the
constraints listed in Sec. III, and the argument that the NMSSM
should keep perturbative up to the Planck scale requires $\lambda $
and $\kappa$ to be smaller than 0.7.

The differences of the NMSSM and MSSM  come from the Higgs sector
and the neutralino sector. In the Higgs sector, now we have three
CP-even and two CP-odd Higgs bosons.  In the basis
$[Re(H_u^0),Re(H_d^0), Re(S)]$, the mass-squared matrix entries for CP-even Higgs
bosons are \cite{Franke,Ellwanger}
\begin{eqnarray}
{\cal M}_{S,11}^2 & = & m_A^2 \cos^2 \beta + m_Z^2 \sin^2 \beta,  \nonumber \\
{\cal M}_{S,22}^2 & = & m_A^2 \sin^2 \beta + m_Z^2 \cos^2 \beta, \nonumber \\
{\cal M}_{S,33}^2 & = & \frac{\lambda^2 v^2}{4 \mu^2} m_A^2 \sin^2 2 \beta
  -\frac{\lambda \kappa}{2} v^2 \sin 2 \beta
  + \frac{\kappa}{\lambda^2} \mu ( \lambda A_\kappa + 4 \kappa  \mu ), \nonumber \\
{\cal M}_{S,12}^2 & = & (2 \lambda^2 v^2 - m_Z^2 - m_A^2 ) \sin \beta \cos \beta, \nonumber  \\
{\cal M}_{S,13}^2 & = & 2 \lambda \mu v \sin \beta
  - \frac{\lambda v}{2 \mu} m_A^2 \sin 2 \beta \cos \beta - \kappa \mu v \cos \beta, \nonumber \\
{\cal M}_{S,23}^2 & = & 2 \lambda \mu v \cos \beta
   - \frac{\lambda v}{2 \mu} m_A^2 \sin \beta \sin 2 \beta - \kappa \mu v \sin \beta,
\label{CP-even}
\end{eqnarray}
and for the CP-odd Higgs bosons, their mass-squared matrix entries
in the basis $[\tilde{A}, Im(S)]$ with
$\tilde{A} = \cos \beta~ Im(H_u^0) + \sin \beta~ Im(H_d^0)$ are
\begin{eqnarray}
{\cal M}_{P,11}^2 & = & \frac{2
\mu}{\sin 2 \beta } ( A_\lambda + \frac{\kappa \mu}{\lambda} ) \equiv m_A^2,  \nonumber \\
{\cal M}_{P,22}^2 & = & \frac{3}{2} \lambda \kappa v^2 \sin 2 \beta
+ \frac{\lambda^2 v^2}{4 \mu^2} m_A^2 \sin^2 2 \beta  - 3 \frac{\kappa}{\lambda} \mu A_\kappa, \nonumber \\
{\cal M}_{P,12}^2 & = & \frac{\lambda v}{2 \mu} m_A^2 \sin 2 \beta
  - 3 \kappa \mu v.   \label{CP-odd}
\end{eqnarray}
Eqs.(\ref{CP-even}) and (\ref{CP-odd}) indicate that the parameters
$\lambda $ and $\kappa  \mu $ affect the mixings of the doublet fields
with the singlet field, $A_\kappa$ only affects the squared-mass of the
singlet field, and in the limit $\lambda, \kappa  \to 0 $, the NMSSM can
recover the MSSM.
One can also learn that in case of small $\lambda$ and $\kappa$
so that the mixings are small, the physical state
with the singlet being the dominant component should couple weakly
to bottom quarks and thus its loop contribution to
$\rho_b$ and $\sin^2 \theta_{eff}^b$ should be small.

The NMSSM predicts five neutralinos, and in the basis
$(- i\lambda_1, - i \lambda_2, \psi_u^0, \psi_d^0, \psi_s )$
their mass matrix is given by \cite{Franke,Ellwanger}
\begin{eqnarray}
\left( \begin{array}{ccccc}
M_1 & 0 & m_Z \sin \theta_W \sin \beta & - m_Z \sin \theta_W \cos \beta  & 0 \\
& M_2 & -m_Z \cos \theta_W \sin \beta & m_Z \cos \theta_W \cos \beta  & 0 \\
& & 0 & -\mu & -\lambda v \cos \beta \\
& & & 0 & - \lambda v \sin \beta \\
& & & & 2 \frac{\kappa}{\lambda} \mu \end{array} \right) . \label{mass
matrix}
\end{eqnarray}
This mass matrix is independent of $A_\kappa$, and the role of $\lambda$
is to introduce the mixings of $\psi_s $ with  $\psi_u^0 $ and
$\psi_d^0$, and $k \mu $ is to affect the mass of $\psi_s$. Quite
similar to the discussion about the Higgs bosons, in case of
small $\lambda$, the correction to $\rho_b$ and $\sin^2
\theta_{eff}^b$ should be insensitive to the value of $\kappa \mu$.

\subsection{NMSSM correction to $\rho_b $ and $\sin^2 \theta_{eff}^b$}
We first look at the SUSY-EW corrections in the NMSSM.
Compared with the corresponding MSSM corrections, the NMSSM effects
involve two additional parameters $\lambda$ and $\kappa$.
As discussed below Eq.(\ref{mass matrix}), in case of small $\lambda$,
the corrections are insensitive to $\kappa$ (our numerical results
verified this conclusion), and thus here we mainly study the dependence
on $\lambda$. We choose a value for $\kappa$ so that the allowed range
of $\lambda$ is wide.
%%%%%%%%%%%%%%%%%%%%%%%%%%%%%%%%%%%%%%%%%%%%%%%%%%%%%%%%%%%%%%%%%%%%
\begin{figure}[tbp]
\epsfig{file=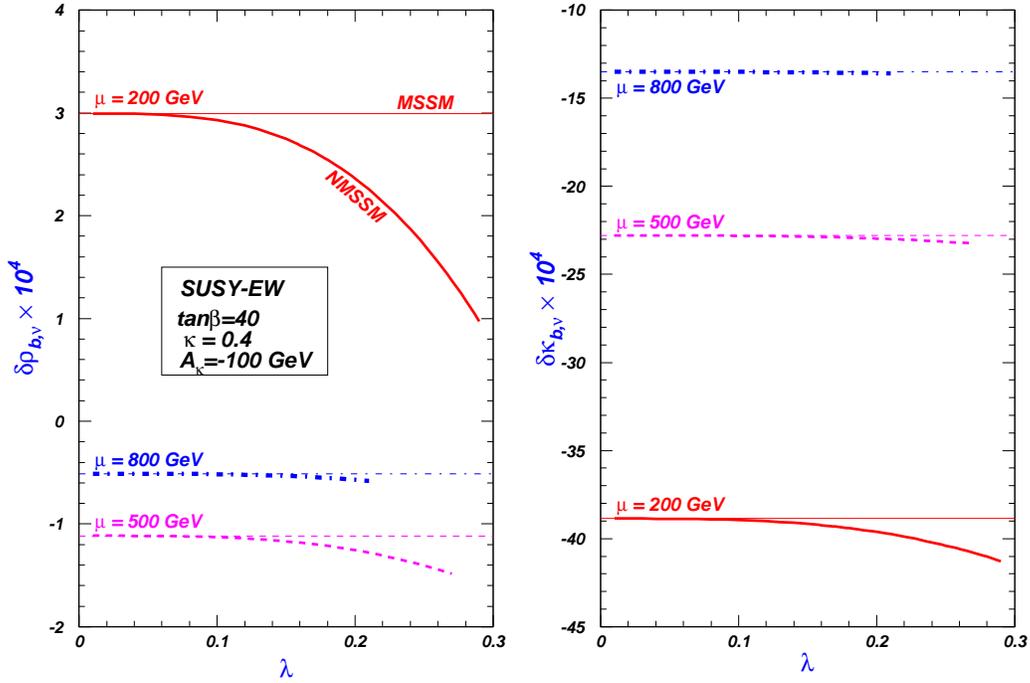,width=14cm}
\vspace{-0.7cm}
\caption{The NMSSM electroweak contributions to $\delta \rho_{b, v}$ and
 $\delta \kappa_{b,v}$, compared with the corresponding MSSM contributions (
 thin horizontal lines), under constraints (1-5).}
\label{NMSSM-EW}
\end{figure}
%%%%%%%%%%%%%%%%%%%%%%%%%%%%%%%%%%%%%%%%%%%%%%%%%%%%%%%%%%%%%%%%%%%%

In Fig.\ref{NMSSM-EW} we show the SUSY-EW contributions to $\delta
\rho_{b,v}$ and $\delta \kappa_{b,v}$ as a function of $\lambda$, in
which $\tan \beta = 40$, $\kappa=0.4$, $A_\kappa = -100$ GeV and
other parameters are same as in Fig.\ref{SUSY-EW1}. One character of
this figure is that both $\delta \rho_{b,v}$ and $\delta
\kappa_{b,v}$ become more negative with the increase of $\lambda$,
which enlarges the gap between the theoretical values and the
experimental data. Another character of this figure is that the
contributions are less sensitive to $\lambda$ when $\mu$ becomes
large. This can be explained from Eq.(\ref{mass matrix}) which shows
that the mixings between $\psi_s $ and the doublets $(\psi_u^0,
\psi_d^0)$ become negligiblly small for sufficiently large $\mu$ and
thus reduce the sensitivity of the contributions to $\lambda$.

%%%%%%%%%%%%%%%%%%%%%%%%%%%%%%%%%%%%%%%%%%%%%%%%%%%%%%%%%%%%%%%%%%%%
\begin{figure}[tbp]
\epsfig{file=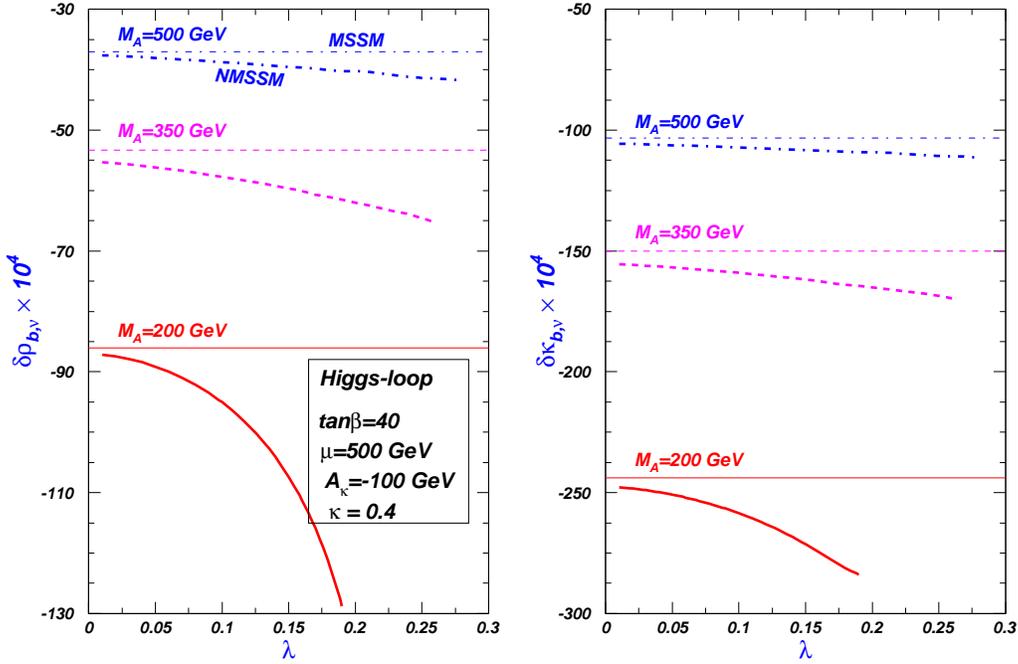,width=14cm}
\vspace{-0.7cm}
\caption{Same as Fig. \ref{NMSSM-EW}, but for the Higgs loop contributions.}
\label{NMSSM-Higgs1}
\end{figure}
%%%%%%%%%%%%%%%%%%%%%%%%%%%%%%%%%%%%%%%%%%%%%%%%%%%%%%%%%%%%%%%%%%%%
%%%%%%%%%%%%%%%%%%%%%%%%%%%%%%%%%%%%%%%%%%%%%%%%%%%%%%%%%%%%%%%%%%%%
\begin{figure}[tbp]
\epsfig{file=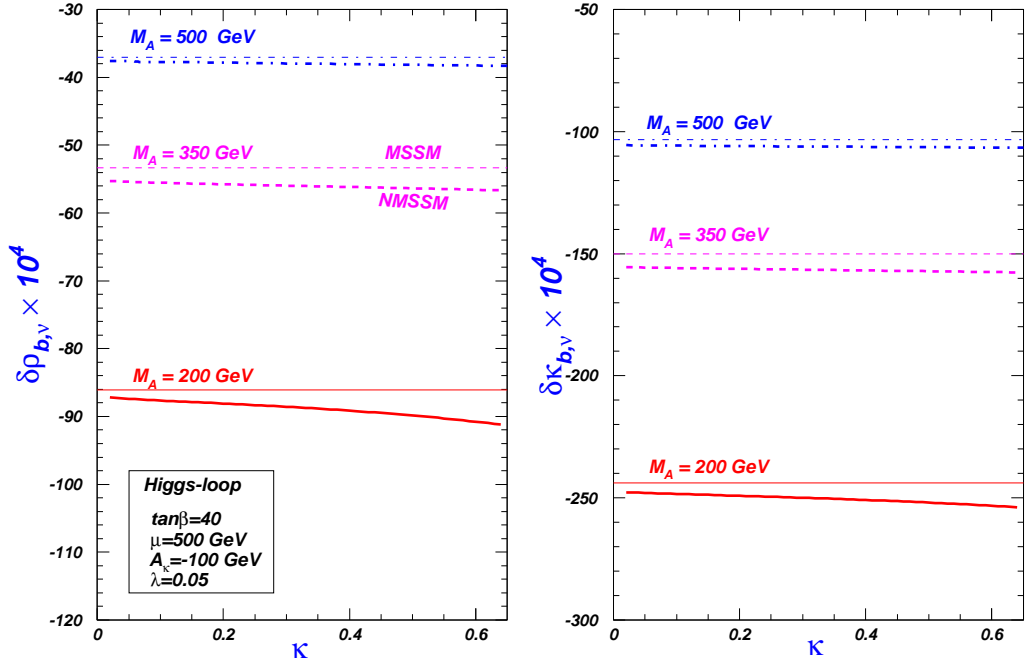,width=14cm}
\vspace{-0.5cm}
\caption{Same as Fig. \ref{NMSSM-EW}, but for the Higgs loop contributions
versus the parameter $\kappa$.}
\label{NMSSM-Higgs2}
\end{figure}
%%%%%%%%%%%%%%%%%%%%%%%%%%%%%%%%%%%%%%%%%%%%%%%%%%%%%%%%%%%%%%%%%%%%

We now turn to the Higgs loop contributions to $\delta \rho_{b,v} $
and $\delta \kappa_{b,v}$ in the NMSSM. For these contributions,
besides $m_A$ and $\tan \beta$, the parameters $\lambda $, $\kappa$
and $A_\kappa$ are also involved. Noting that these contributions
are more sensitive to $\lambda$ and $\kappa$ than to $A_\kappa$, we
only study their dependence on $\lambda$ and $\kappa$.

In Fig.\ref{NMSSM-Higgs1} we show the contributions versus $\lambda $,
where $\tan \beta = 40 $, $ \kappa = 0.4 $, $A_\kappa = - 100$ GeV
and other parameters are same as in Fig.\ref{SUSY-Higgs}.
This figure shows the same behavior as in Fig.\ref{NMSSM-EW},
and the dependence on $\lambda$ becomes rather weak in case of large $m_A$.

In Fig.\ref{NMSSM-Higgs2}, we show the dependence of the
contributions on $\kappa$, as shown. This figure exhibits the
similar behavior to Fig.\ref{NMSSM-Higgs1}. Compared with
Fig.\ref{NMSSM-Higgs1} and Fig.\ref{NMSSM-Higgs2}, one can learn
that the contributions have a stronger dependence on $\lambda $ than
on $\kappa$.

%%%%%%%%%%%%%%%%%%%%%%%%%%%%%%%%%%%%%%%%%%%%%%%%%%%%%%%%%%%%%%%%%%%%
\begin{figure}[tbp]
\epsfig{file=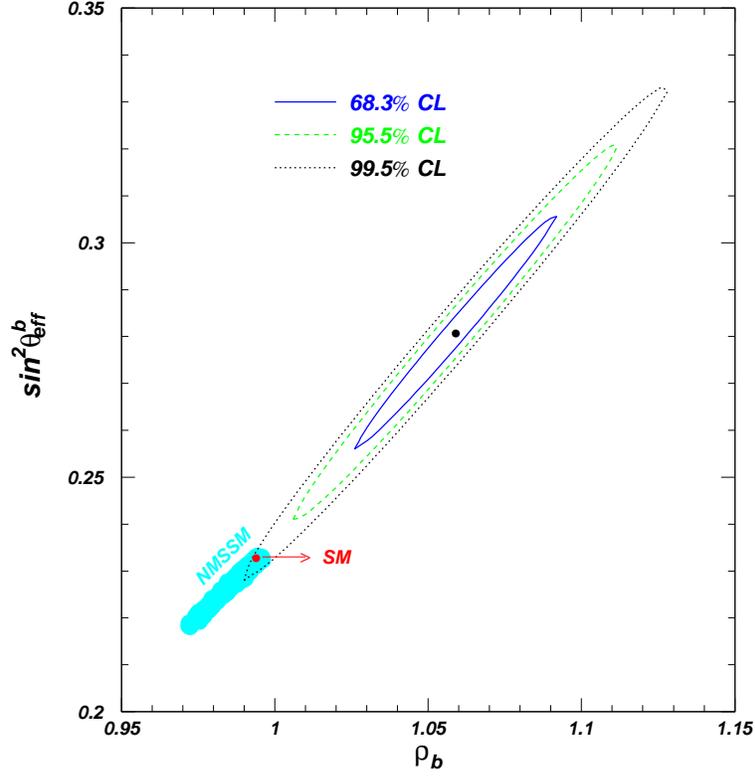,width=10cm} \vspace{-0.5cm}
\caption{\small Same as Fig. \ref{contours}, but for the NMSSM predictions.}
\label{contours1}
\end{figure}
%%%%%%%%%%%%%%%%%%%%%%%%%%%%%%%%%%%%%%%%%%%%%%%%%%%%%%%%%%%%%%%%%%%%

Like in Fig.\ref{contours}, we also investigate the extent to
which the NMSSM predictions can agree with the experiment by
scanning over the SUSY parameter space in the region of
Eq.(\ref{region}) and
\begin{eqnarray}
\lambda, \kappa \leq 0.7, \quad \quad -1 {\rm~ TeV} < A_\kappa < 1 {\rm~ TeV}.
\end{eqnarray}
Our result is shown in Fig.\ref{contours1}. Compared with
Fig.\ref{contours}, one can learn that the NMSSM cannot
improve the agreement and instead may exacerbate
the agreement in a large part of the allowed parameter space.

If we define a quantity $F(\lambda, \kappa) - F(0, 0 )$ with $F$
denoting either $\delta \rho_{b,v}$ or $\delta \kappa_{b,v}$ with
$F(\lambda, \kappa) $ being the value of $F$ in the NMSSM with
arbitrary values of $\lambda $ and $\kappa$, and $F(0,0)$ being the
value of $F$ in the MSSM limit, then by studying various cases we
find this quantity is generally smaller than $5 \times 10^{-3}$,
which means that in the allowed region for $\lambda$ and $\kappa$,
NMSSM only slightly modifies the MSSM predictions of $\rho_b$ and
$\sin^2 \theta_{eff}^b$.

\section{Conclusions}
The $Z b \bar{b}$ coupling determined from the $Z$-pole
measurements at LEP/SLD deviate significantly from the SM
prediction. In terms of $\rho_b $ and $\sin^2 \theta_{eff}^b$, the
SM prediction is about $3\sigma$ below the experimental data. If
this anomaly is not a statistical or systematic effect, it would
signal the presence of new physics in association with the $Zb\bar
b$ coupling. In this work we scrutinized the full one-loop
supersymmetric effects on $Z b \bar{b}$ coupling in both the MSSM
and the NMSSM, considering all current constraints which are from
the precision electroweak measurements, the direct search for
sparticles and Higgs bosons, the stability of Higgs potential, the
dark matter relic density, and the muon g-2 measurement. We
analyzed the characters of each type of the corrections and
searched for the SUSY parameter regions where the corrections
could be sizable. We found that the potentially sizable
corrections come from the Higgs sector with light $m_A$ and large
$\tan \beta$, which can reach $-2\%$ and $-6\%$ for $\rho_b $ and
$\sin^2 \theta_{eff}^b$, respectively. However, such sizable negative
corrections  are just opposite to what needed to
solve the anomaly. We also scanned over the allowed parameter
space and investigated to what extent supersymmetry can narrow the
discrepancy between theoretical predictions and the experimental
values. We found that under all current constraints, the
supersymmetric effects are quite restrained and cannot
significantly ameliorate the anomaly  of $Zb\bar b$ coupling.
Compared with $\chi^2/dof = 9.62/2$ in the SM,  the MSSM and NMSSM
can only improve it to $\chi^2/dof = 8.77/2$ in the allowed
parameter space.

In the future the GigaZ option at the proposed
International Linear Collider (ILC) with an integrated luminosity of
30 fb$^{-1}$ is expected to produce more than $10^9$ $Z$-bosons
\cite{ILC} and will give a more precise measurement of
$Z b \bar{b}$ coupling, which will allow for a test of new physics models.
If the anomaly of $Z b \bar{b}$ coupling persists,
it would suggest new physics beyond the MSSM and NMSSM.
One possible form of such new physics is the model with
additional right-handed gauge bosons which couple predominantly to
the third generation quarks \cite{He}. These new gauge bosons usually
mix with $Z$ and $W$ so that the $Z b_R \bar{b}_R $ and $W b_R
\bar{t}_R$ couplings in the SM may be greatly changed.  A careful
investigation of top quark processes at the LHC, such as top quark decay
to the polarized W boson \cite{Fischer}, may test this model in the near
future.

\section*{Acknowledgement}
This work was supported in part by the National Sciences and
Engineering Research Council of Canada, by the National Natural
Science Foundation of China (NNSFC) under grant No. 10505007,
10725526 and 10635030, and by HASTIT under grant No. 2009HASTIT004.

\appendix

\section{gauge boson self-energy in NMSSM}

In the NMSSM the contributions to vector boson self-energy come from
the loops mediated by the SM fermions, gauge bosons, Higgs bosons,
sfermions, charginos and neutralinos, respetively. In the following
we list the expressions for pure new physics contributions, namely
from the loops of Higgs bosons, sfermions, charginos and
neutralinos, respectively. We adopt the convention of
 \cite{Ellwanger} for the SUSY parameters.

\begin{itemize}

\item[(1)] Higgs contribution:

The NMSSM has an extended Higgs boson sector with a pair of charged
Higgs bosons $H^\pm$, two CP-odd Higgs boson $a_i$ and three CP-even
Higgs boson $h_i$. The Higgs contribution to gauge boson self-energy
arises from $VHH$, $VVHH$ and $VVH$ interactions and because we
choose 't Hooft-Feynman gauge to calculate the contribution, the
gauge boson contribution and the Higgs contribution are in general
entangled. In our calculation, we are actually interested in the
difference between the contribution from the NMSSM Higgs sector and
that from the SM Higgs sector (see the discussion in the last
paragraph of Sect. II). Since the SM contribution is well
known\cite{Hollik,Denner}, we only list the NMSSM contribution.
\begin{eqnarray}
\Sigma^{T}_{\gamma \gamma} (p^2) &=& \frac{e^2}{16 \pi^2} B_5 (p, m_{H^+}, m_{H^+}), \\
\Sigma^{T}_{\gamma Z} (p^2) &=& \frac{1}{16 \pi^2} \frac{e g \cos 2
\theta_W}{2 \cos \theta_W}  B_5 (p, m_{H^+}, m_{H^+}),
\end{eqnarray}
\begin{eqnarray}
\Sigma^{T}_{ZZ} (p^2) &=& \frac{1}{16 \pi^2} \frac{g^2}{4 \cos^2
\theta_W} \biggl\{ \biggl[ ( |S_{i1}|^2 + |S_{i2}|^2 ) A(m_{h_i}) +
|P_{i1}^\prime|^2 A(m_{a_i}) + A(m_Z )
 \nonumber \\
& & - 4  | \sin
\beta  S_{i2} -\cos \beta S_{i1} |^2 | P_{j1}^\prime|^2 B_{22}(p, m_{a_j}, m_{h_i})  \nonumber \\
& & - 4 | \cos
\beta  S_{i2} + \sin \beta S_{i1} |^2 B_{22}(p, m_{Z}, m_{h_i}) \biggr]  \nonumber \\
&& +2 \cos^2 2 \theta_W  \biggl[ A(m_{H^+}) - 2
B_{22} (p, m_{H^+}, m_{H^+} ) \biggr ]  \nonumber \\
&&  + 4 m_Z^2 |\cos\beta S_{i2} + \sin \beta S_{i1}|^2 B_0 (p, m_Z,
m_{h_i}) \biggr\},\\
\Sigma^{T}_{WW} (p^2) &=& \frac{1}{16 \pi^2} \frac{g^2}{4} \biggl\{
  \biggl[ A(m_{H^+}) + ( |S_{i1}|^2 + |S_{i2}|^2 ) A(m_{h_i}) + A(m_W) \nonumber \\
& &  - 4 |\sin \beta  S_{i2} -\cos \beta S_{i1}|^2 B_{22}(p, m_{H^+}, m_{h_i})  \nonumber \\
& & - 4 |\cos \beta  S_{i2} + \sin \beta S_{i1}|^2 B_{22}(p, m_{W}, m_{h_i}) \biggr]\nonumber \\
&&  + \biggl[ A(m_{H^+}) + | P_{i1}^\prime |^2 A(m_{a_i})
    - 4 | P_{i1}^\prime |^2 B_{22} (p, m_{H^+}, m_{a_i} ) \biggr ]  \nonumber \\
&&  + 4 m_W^2 |\cos\beta S_{i2} + \sin \beta S_{i1} |^2 B_0 (p, m_W,
m_{h_i} ) \biggr\},
\end{eqnarray}
In above equations, $g$ is the SU(2) gauge coupling, and $S$ and
$P^\prime$ are the rotation mass matrices defined in the Appendix A
of \cite{Ellwanger} to diagonalize CP-even and CP-odd Higgs mass
matrices, respectively. $A$ and $B_{22}$ are the standard one- and
two-point loop functions firstly defined in \cite{Passarino}. $B_5$
is related with standard loop functions by \cite{Hagiwara}
\begin{eqnarray}
B_5(p, m_1, m_2) = A(m_1) + A(m_2) - 4 B_{22}(p, m_1, m_2).
\end{eqnarray}

\item[(2)] Sfermion contribution:

The sfermion contributions are given by
\small
\begin{eqnarray}
\Sigma^T_{WW} (p^2) &=& \frac{1}{16 \pi^2} \frac{g^2}{2} C_f
R^{\tilde{u} \ast}_{\alpha 1} R^{\tilde{u}}_{\alpha 1} R^{\tilde{d}
\ast}_{\beta 1} R^{\tilde{d}}_{\beta 1} B_5(p, m_{\tilde{u}_\alpha},
m_{\tilde{d}_\beta} ) ,
\end{eqnarray}
\begin{eqnarray}
\Sigma^T_{ZZ} (p^2) &=& \frac{1}{16 \pi^2}
   \frac{g^2}{\cos^2 \theta_W} C_f
   \biggl\{ I_{3f}^2 R^{\tilde{f} \ast}_{\alpha 1}
   R^{\tilde{f}}_{\alpha 1} R^{\tilde{f} \ast}_{\beta 1}
   R^{\tilde{f}}_{\beta 1} B_5(p, m_{\tilde{f}_\alpha}, m_{\tilde{f}_\beta} )  \nonumber  \\
&& - 2 s_W^2 I_{3f} Q_f R^{\tilde{f} \ast}_{\alpha 1}
   R^{\tilde{f}}_{\alpha 1} B_5(p, m_{\tilde{f}_\alpha},
   m_{\tilde{f}_\alpha} ) + s_W^4 Q_f^2 B_5(p,
   m_{\tilde{f}_\alpha}, m_{\tilde{f}_\alpha} )  \biggr\}, \\
\Sigma^T_{\gamma \gamma} (p^2) &=& \frac{e^2}{16 \pi^2} C_f Q_f^2
    B_5(p, m_{\tilde{f}_\alpha}, m_{\tilde{f}_\alpha} ),\\
\Sigma^T_{\gamma Z} (p^2) &=& \frac{e}{16 \pi^2} \frac{g}{\cos\theta_W}
    C_f \biggl\{ I_{3f} Q_f R^{\tilde{f} \ast}_{\alpha 1}
    R^{\tilde{f}}_{\alpha 1} -Q_f^2 s_W^2  \biggr\}
    B_5(p, m_{\tilde{f}_\alpha}, m_{\tilde{f}_\alpha} ),
\end{eqnarray}
\normalsize
where the color factor $C_f$ is 3 for squarks and 1 for sleptons.
The electric charge $Q_f$ is given by $2/3 ,-1/3, 0, -1$ for
$\tilde{u}, \tilde{d}, \tilde{\nu}_l, \tilde{l}$, respectively.
$I_{3f}$ denotes the third component of the weak isospin, which is
$+1/2$ and $-1/2$ for the up- and down-type sfermions, respectively.
$R$ is the rotation matrix to diagonalize sfermion mass matrix.

\item[(3)] Chargino and neutralino contribution:

For a generic interaction between a vector boson and two fermions,
it contributes to vector boson self-energy in the form:
\begin{eqnarray}
\Sigma_{V^\prime V}^T (p^2) &= & \frac{2}{16 \pi^2} \biggl\{ (
g_L^{\bar{\psi}_j \psi_i V^\prime } g_L^{\bar{\psi}_i \psi_j V^\ast}
+ g_R^{\bar{\psi}_j \psi_i V^\prime } g_R^{\bar{\psi}_i \psi_j
V^\ast} ) ( 2 p^2 B_3 - B_4 )(p, m_{\psi_i}, m_{\psi_j} )  \nonumber
\\ && +  ( g_L^{\bar{\psi}_j \psi_i V^\prime } g_R^{\bar{\psi}_i \psi_j
V^\ast}  + g_R^{\bar{\psi}_j \psi_i V^\prime} g_L^{\bar{\psi}_i
\psi_j V^\ast} ) m_{\psi_i} m_{\psi_j} B_0 (p, m_{\psi_i},
m_{\psi_j} ) \biggr\},  \label{fermion}
\end{eqnarray}
where $g_{L,R}^{\bar{\psi}_i \psi_j V} $ is the coupling strength of the
vector boson with left-handed or righ-handed fermions.
The functions $B_3$ and $B_4$ are related with the standard two-point functions
by \cite{Hagiwara}
\begin{eqnarray*}
B_3(p, m_1, m_2) &=& - B_1(p, m_1, m_2) - B_{21}(p, m_1, m_2), \\
B_4(p, m_1, m_2) &=& - m_1^2 B_1(p, m_2, m_1) - m_2^2 B_1(p, m_1, m_2).
\end{eqnarray*}
For the charginos and neutralinos, the coefficients of their
interactions with vector bosons take following forms:
\small
\begin{eqnarray*}
&& g_L^{\bar{\tilde{\chi}}^0_i \tilde{\chi}^+_j W^-} = g ( -
   \frac{1}{\sqrt{2}} N_{i3} V_{j2}^\ast + N_{i2} V_{j1}^\ast ), \quad
   g_R^{\bar{\tilde{\chi}}^0_i \tilde{\chi}^+_j W^-} = g (
   \frac{1}{\sqrt{2}} N_{i4}^\ast U_{j2} + N_{i2}^\ast U_{j1} ), \\
&& g_L^{\bar{\tilde{\chi}}^0_i \tilde{\chi}^0_j Z} =
  \frac{g}{2 \cos\theta_W} ( -  N_{i4} N_{j4}^\ast
  +  N_{i3} N_{j3}^\ast ), \quad
  g_R^{\bar{\tilde{\chi}}^0_i \tilde{\chi}^0_j Z} =
  \frac{g}{2\cos\theta_W} (  N_{i4}^\ast N_{j4} -  N_{i3}^\ast N_{j3} ), \\
&& g_L^{\bar{\tilde{\chi}}^+_i \tilde{\chi}^+_j Z}
   = \frac{g}{\cos\theta_W} ( - V_{i1} V_{j1}^\ast
     - \frac{1}{2} V_{i2} V_{j2}^\ast + \delta_{ij} \sin^2 \theta_W ), \quad
   g_L^{\bar{\tilde{\chi}}^+_i \tilde{\chi}^+_j \gamma} = - e \delta_{ij}, \\
&& g_R^{\bar{\tilde{\chi}}^+_i \tilde{\chi}^+_j Z} =
   \frac{g}{\cos\theta_W} ( - U_{i1}^\ast U_{j1} - \frac{1}{2}
   U_{i2}^\ast U_{j2} + \delta_{ij} \sin^2 \theta_W ), \quad
   g_R^{\bar{\tilde{\chi}}^+_i \tilde{\chi}^+_j \gamma} = - e \delta_{ij}.
\end{eqnarray*}
\normalsize
But as for the contribution from neutralino sector, one should note
that, due to the Majorana nature of neutralinos, an addition factor
$\frac{1}{2}$ should be multiplied when using above formulae to get
neutralino contribution to $Z$-boson self-energy.
\end{itemize}

\section{Vertex corrections to $Z\to f \bar{f}$ in NMSSM}

In this section we present the expressions of the radiative
correction to $Z \bar{f} f$ vertex in the NMSSM, namely $\delta v_f
$ and $ \delta a_f$ defined in Eq.(\ref{original}). In our
calculation we neglect terms proportional to fermion mass except
for $f= b $ (bottom quark) where we keep terms proportional to
bottom quark Yukawa coupling, $Y_b \sim \frac{m_b}{\cos \beta}$,
since those terms may be enhanced by large $\tan \beta$.  Throughout
this section all $Z$-boson coupling coefficients, such as $\delta v_f
$ and $\delta a_f$,  are defined so that the common factor
$e/(2 \sin \theta_W \cos \theta_W )$ has been extracted.

To neatly present $\delta v_f$ and $\delta a_f$, it is convenient to
introduce the quantities $\delta g_{\lambda}^f$ with $\lambda =L,
R$, which denote the vertex correction to $Z \bar{f}_\lambda
f_\lambda $ interaction and are related with $\delta v_f$ and
$\delta a_f$ by $\delta v_f = ( \delta g_L^f + \delta g_R^f )/2 $
and $\delta a_f = ( \delta g_L^f - \delta g_R^f )/2$, respectively.
$\delta g_\lambda^f$ is given by \cite{Hollik}
\begin{eqnarray}
 \delta g_\lambda^f &=& \Gamma_{f_\lambda}(m_Z^2) - g_\lambda^{Z \bar{f}f} \Sigma_{f_\lambda}
 (m_f^2) - 2 \delta_{\lambda L} a_f \frac{\cos \theta_W}{\sin \theta_W}
 \frac{\Sigma^{\gamma Z} (0)}{m_Z^2}, \label{vertex correction}
\end{eqnarray}
where $\Gamma_{f_\lambda}$ is the unrenormalized vertex correction
to $Z \bar{f}_\lambda f_\lambda$ interaction,  the second term on
the RHS  denotes the counter term arising from the fermion
$f_\lambda$ self-energy, and the last term is the counter term from
the vector boson self-energy.

Assuming the interaction between scalars $\phi_i$ with $Z $ boson
takes the form $\Gamma^{\phi^\ast_i \phi_j Z} = g^{\phi^\ast_i
\phi_j Z} ( p_{\phi_i} + p_{\phi_j} ) $, we can write down
$\Sigma_{f_\lambda}(m_f^2)$ and the vertex function
$\Gamma_{f_\lambda}(q^2)$ mediated by a fermion $\psi$ and a scalar
$\phi$ in a compact generic notation as
\begin{eqnarray}
(4\pi)^2 \Sigma_{f_\lambda}(p_f^2) &=&
    C_g \biggl| g^{\bar{\psi}_j f \phi_i^\ast}_\lambda \biggr|^2
    \biggl( B_0 + B_1 \biggr) (p_f, m_{\phi_i}, m_{\psi_j}), \\
 (4\pi)^2 \Gamma_{f_\lambda}(q^2) &=&
    - C_g \Biggl\{
    \biggl( g_\lambda^{\bar{\psi}_j f \phi_k^\ast} \biggr)^*
    g_\lambda^{\bar{\psi}_i f \phi_k^\ast}
    \biggl[
    g_\lambda^{\bar{\psi}_j \psi_i Z} m_{\psi_i} m_{\psi_j} C_0
\nonumber \\
&&
    + g_{-\lambda}^{\bar{\psi}_j \psi_i Z}
    \biggl\{-q^2 (C_{12} + C_{23}) - 2 C_{24} +  \frac{1}{2} \biggr\}
    \biggr] (p_{\bar{f}}, p_{f}, m_{\psi_i}, m_{\phi_k}, m_{\psi_j})
\nonumber \\
&&
    -  \biggl( g^{\bar{\psi}_k f \phi_i^\ast}_\lambda \biggr)^*
    g^{\bar{\psi}_k f \phi_j^\ast}_\lambda
    g^{\phi_i^\ast \phi_j Z}
    2 C_{24}(p_{\bar{f}}, p_f, m_{\phi_j}, m_{\psi_k}, m_{\phi_i})
    \Biggr\}.
\end{eqnarray}
Here $C_g$ is $4/3$ for the gluino contribution ($\psi=gluino$) and
1 for the others. The chirality index $-\lambda$ follows the rule:
$-L=R, -R=L$.

If $f$ is a lepton, the following combination of $\{ \psi, \phi
\}$ contribute to the vertex:
\begin{itemize}
\item{Chargino correction}:
\begin{eqnarray}
&& \{\psi , \phi \} =\{ \tilde{\chi}^- ,  \tilde{\nu} \}:
\nonumber \\
&& g_L^{\bar{\tilde{\chi}}^-_j l \tilde{\nu}^\ast} = - g
V_{j1}^\ast; \quad \quad  g_R^{\bar{\tilde{\chi}}^-_j l
\tilde{\nu}^\ast} = 0 ; \quad \quad  g^{\tilde{\nu}^\ast \tilde{\nu}
Z} = - 1 ; \nonumber \\
&& g_L^{\bar{\tilde{\chi}}^-_j \tilde{\chi}^-_i Z} = 2 ( U_{i1}^\ast
U_{j1} + \frac{1}{2}
U_{i2}^\ast U_{j2} - \delta_{ij} \sin^2 \theta_W ); \nonumber \\
&&g_R^{\bar{\tilde{\chi}}^-_j \tilde{\chi}^-_i Z} = 2 ( V_{i1}
V_{j1}^\ast + \frac{1}{2} V_{i2} V_{j2}^\ast - \delta_{ij} \sin^2
\theta_W );
\end{eqnarray}
\item{Neutralino correction}:
\begin{eqnarray}
&& \{\psi , \phi \} =\{ \tilde{\chi}^0 ,  \tilde{l} \}:
\nonumber \\
&& g_L^{\bar{\tilde{\chi}}^0_j l \tilde{l}_\alpha^\ast} =
\frac{g}{\sqrt{2}} R^{\tilde{l}}_{\alpha 1} ( N_{j2}^\ast + \tan
\theta_W N_{j1}^\ast ) ; \quad \quad  g_R^{\bar{\tilde{\chi}}^0_j l
\tilde{l}_\alpha^\ast} = - \sqrt{2} g R^{\tilde{l}}_{\alpha 2} \tan
\theta_W N_{j1} ;
\nonumber \\
&& g_L^{\bar{\tilde{\chi}}^0_j \tilde{\chi}^0_i Z} = - N_{j4}
N_{i4}^\ast +  N_{j3} N_{i3}^\ast ; \quad \quad
g_R^{\bar{\tilde{\chi}}^0_j \tilde{\chi}^0_i Z} = N_{j4}^\ast N_{i4} -  N_{j3}^\ast N_{i3};  \nonumber \\
&&g^{\tilde{l}_\alpha^\ast \tilde{l}_\beta Z} = ( 1 - 2 \sin^2
\theta_W ) R^{\tilde{l}}_{\alpha 1} R^{\tilde{l} \ast}_{\beta 1} - 2
\sin^2 \theta_W R^{\tilde{l}}_{\alpha 2} R^{\tilde{l} \ast}_{\beta
2};
\end{eqnarray}
\end{itemize}

If $f$ is the bottom quark, the following combination of $\{ \psi,
\phi \}$ contribute to the vertex:
\begin{itemize}
\item{Chargino correction}:
\begin{eqnarray}
&& \{\psi , \phi \} =\{ \tilde{\chi}^- ,  \tilde{t} \}:
\nonumber \\
&& g_L^{\bar{\tilde{\chi}}^-_j b \tilde{t}_\alpha^\ast} = g ( -
R^{\tilde{t}}_{\alpha 1} V_{j1}^\ast + Y_t R^{\tilde{t}}_{\alpha 2}
V_{j2}^\ast ); \quad g_R^{\bar{\tilde{\chi}}^-_j b
\tilde{t}_\alpha^\ast} = g R^{\tilde{t}}_{\alpha 1} Y_b U_{j2};
\nonumber \\
&&g^{\tilde{t}_\alpha^\ast \tilde{t}_\beta Z} = ( -1 + \frac{4}{3}
\sin^2 \theta_W ) R^{\tilde{t}}_{\alpha 1} R^{\tilde{t} \ast}_{\beta
1} + \frac{4}{3} \sin^2 \theta_W R^{\tilde{t}}_{\alpha 2}
R^{\tilde{t} \ast}_{\beta 2};
\end{eqnarray}
Note that in order to write the couplings in a neat form, we
define $Y_t = m_t/\sqrt{2} m_W \sin \beta$, and $Y_b = m_b/\sqrt{2}
m_W \cos \beta$. Such definitions differ from their conventional
definitions by a factor $g$. We adopt such a convention throughout
our paper.

\item{Neutralino correction}:
\begin{eqnarray}
&& \{\psi , \phi \} =\{ \tilde{\chi}^0 ,  \tilde{b} \}:
\nonumber \\
&& g_L^{\bar{\tilde{\chi}}^0_j b \tilde{b}_\alpha^\ast} = g \biggl(
\frac{\sqrt{2}}{2} R^{\tilde{b}}_{\alpha 1} ( N_{j2}^\ast -
\frac{1}{3} \tan \theta_W N_{j1}^\ast ) -   Y_b
R^{\tilde{b}}_{\alpha 2} N_{j4}^\ast
\biggr); \nonumber \\
&& g_R^{\bar{\tilde{\chi}}^0_j b \tilde{b}_\alpha^\ast} = - g (
R^{\tilde{b}}_{\alpha 1} Y_b N_{j4} + \frac{\sqrt{2}}{3}
R^{\tilde{b}}_{\alpha 2} \tan \theta_W N_{j1} );
\nonumber \\
&&g^{\tilde{b}_\alpha^\ast \tilde{b}_\beta Z} = ( 1 - \frac{2}{3}
\sin^2 \theta_W ) R^{\tilde{b}}_{\alpha 1} R^{\tilde{b} \ast}_{\beta
1} - \frac{2}{3} \sin^2 \theta_W R^{\tilde{b}}_{\alpha 2}
R^{\tilde{b} \ast}_{\beta 2};
\end{eqnarray}
\item{Gluino correction}:
\begin{eqnarray}
&& \{\psi, \phi \} =\{ \tilde{g} , \tilde{b} \}:
\nonumber \\
&& g_L^{\bar{\tilde{g}} b \tilde{b}_\alpha^\ast} = -\sqrt{2} g_s
R^{\tilde{b}}_{\alpha 1} ; \quad   g_R^{\bar{\tilde{g}} b
\tilde{b}_\alpha^\ast} = \sqrt{2} g_s R^{\tilde{b}}_{\alpha 2} ;
\end{eqnarray}
\item{Charged Higgs contribution}:
\begin{eqnarray}
&& \{\psi , \phi \} =\{ t ,  H^- \}:
\nonumber \\
&& g_L^{\bar{t} b (H^-)^\ast} = \frac{g m_t}{\sqrt{2} m_W} \cot
\beta; \quad
 g_R^{\bar{t} b (H^-)^\ast} = \frac{g m_b}{\sqrt{2} m_W} \tan \beta;
\nonumber \\
&& g_L^{\bar{t}tZ}= - ( 1 - \frac{4}{3} \sin^2 \theta_W ); \quad
g_R^{\bar{t}tZ}=  \frac{4}{3} \sin^2 \theta_W; \nonumber
\\ && g^{(H^-)^\ast H^- Z} = \cos 2
\theta_W
\end{eqnarray}
\item{Neutral Higgs contribution}:
\begin{eqnarray} && \{\psi , \phi \} =\{ b , (h, a, G^0)
\}:
\nonumber \\
&& g_L^{\bar{b} b h_i} = -\frac{g m_b}{2 m_W \cos \beta} S_{i2};
\quad
 g_R^{\bar{b} b h_i} = -\frac{g m_b}{2 m_W \cos \beta} S_{i2};
\nonumber \\
&& g_L^{\bar{b} b a_i} = -\frac{i g m_b}{2 m_W \cos \beta} P_{i2} =
-\frac{i g m_b }{2 m_W } P_{i1}^\prime \tan \beta ;  \nonumber \\
&& g_R^{\bar{b} b a_i} = \frac{i g m_b}{2 m_W \cos \beta} P_{i2} =
\frac{i g m_b}{2 m_W } P_{i1}^\prime \tan \beta;
\nonumber \\
&& g_L^{\bar{b} b G^0} = - \frac{i g m_b}{2 m_W} ; \quad
 g_R^{\bar{b} b G^0} = \frac{i g m_b}{2 m_W};
\nonumber \\
&& g_L^{\bar{b}b Z}= ( 1 - \frac{2}{3} \sin^2 \theta_W ); \quad
g_R^{\bar{b}b Z}= -  \frac{2}{3} \sin^2 \theta_W; \nonumber
\\ && g^{h_i^\ast a_j Z} = - i ( S_{i2}
P_{j2} - S_{i1} P_{j1} ) = - i (S_{i2} \sin \beta - S_{i1} \cos
\beta )
P_{j1}^\prime,  \nonumber \\
&& g^{a_j^\ast h_i Z} =  i ( S_{i2} P_{j2} - S_{i1} P_{j1} ) = i
(S_{i2} \sin \beta
- S_{i1} \cos \beta ) P_{j1}^\prime, \nonumber \\
&& g^{h_i^\ast G^0 Z} = - i  ( S_{i2}
\cos \beta + S_{i1} \sin \beta ),   \nonumber \\
&& g^{G^{0 \ast} h_i Z} = i ( S_{i2} \cos \beta + S_{i1} \sin \beta).
\end{eqnarray}
Note that in the above formulas we did not include the
contribution to $\delta g_\lambda$ from the loop of
$\{t, G^-\}$. Such contribution alone is UV-convergent and should be
attributed to the SM radiative effects. This situation is quite
different for the neutral Higgs contribution where the effects of
the loops of $\{b, G^0\}$ are UV divergence and must be
included with other neutral Higgs contribution to get an finite
result.
\end{itemize}

If $f$ is the charm quark, the following combination of $\{ \psi,
\phi \}$ contribute to the vertex:
\begin{itemize}
\item{Chargino correction}:
\begin{eqnarray}
&& \{\psi , \phi \} =\{ \tilde{\chi}^+ ,  \tilde{s} \}:
\nonumber \\
&& g_L^{\bar{\tilde{\chi}}^+_j c \tilde{s}_\alpha^\ast} = - g
R^{\tilde{s}}_{\alpha 1} U_{j1}^\ast; \quad \quad \quad \quad
g_R^{\bar{\tilde{\chi}}^+_j c \tilde{s}_\alpha^\ast} = 0 ;
\nonumber \\
&& g_L^{\bar{\tilde{\chi}}^+_j \tilde{\chi}^+_i Z} = - 2 (
V_{i1}^\ast V_{j1} + \frac{1}{2}
V_{i2}^\ast V_{j2} - \delta_{ij} \sin^2 \theta_W ); \nonumber \\
&&g_R^{\bar{\tilde{\chi}}^+_j \tilde{\chi}^+_i Z} = - 2 ( U_{i1}
U_{j1}^\ast + \frac{1}{2} U_{i2} U_{j2}^\ast - \delta_{ij} \sin^2
\theta_W ); \nonumber \\
&&g^{\tilde{s}_\alpha^\ast \tilde{s}_\beta Z} = ( 1 - \frac{2}{3}
\sin^2 \theta_W ) R^{\tilde{s}}_{\alpha 1} R^{\tilde{s} \ast}_{\beta
1} - \frac{2}{3} \sin^2 \theta_W R^{\tilde{s}}_{\alpha 2}
R^{\tilde{s} \ast}_{\beta 2};
\end{eqnarray}

\item{Neutralino correction}:
\begin{eqnarray}
&& \{\psi , \phi \} =\{ \tilde{\chi}^0 ,  \tilde{c} \}:
\nonumber \\
&& g_L^{\bar{\tilde{\chi}}^0_j c \tilde{c}_\alpha^\ast} = - \frac{ g
}{\sqrt{2}} R^{\tilde{c}}_{\alpha 1} ( N_{j2}^\ast +
\frac{1}{3} \tan \theta_W N_{j1}^\ast ); \nonumber \\
&& g_R^{\bar{\tilde{\chi}}^0_j c \tilde{c}_\alpha^\ast} = \frac{2
\sqrt{2} g}{3} R^{\tilde{c}}_{\alpha 2} \tan \theta_W N_{j1} ;
\nonumber \\
&&g^{\tilde{c}_\alpha^\ast \tilde{c}_\beta Z} = ( - 1 + \frac{4}{3}
\sin^2 \theta_W ) R^{\tilde{c}}_{\alpha 1} R^{\tilde{c} \ast}_{\beta
1} + \frac{4}{3} \sin^2 \theta_W R^{\tilde{c}}_{\alpha 2}
R^{\tilde{c} \ast}_{\beta 2};
\end{eqnarray}

\item{Gluino correction}:
\begin{eqnarray}
&& \{\psi, \phi \} =\{ \tilde{g} , \tilde{c} \}:
\nonumber \\
&& g_L^{\bar{\tilde{g}} c \tilde{c}_\alpha^\ast} = -\sqrt{2} g_s
R^{\tilde{c}}_{\alpha 1} ; \quad   g_R^{\bar{\tilde{g}} c
\tilde{c}_\alpha^\ast} = \sqrt{2} g_s R^{\tilde{c}}_{\alpha 2} ;
\end{eqnarray}

\end{itemize}

The above expressions then suffice to calculate all the $Z f_\alpha
\bar{f}_\alpha$ vertex corrections $\delta g_\alpha^f$. Summation
should be taken over all non-vanishing coupling combinations, such as
over the indices of sfermions, charginos, neutralinos,
scalar Higgs and pseudo-scalar Higgs.

\section{NMSSM contributions to the $\mu$-decay}

In the NMSSM the flavor-dependent correction to the decay
$\mu \to \nu_\mu e \bar{\nu}_e$ mainly comes from the loops
mediated by gauginos, and the corrected amplitude can be written
as \cite{Heinemeyer}
\begin{eqnarray}
M = M_B \left( 1 + 2 \delta^{(v)} + \delta^{(b)} \right),
\end{eqnarray}
where $M_B$ is the Born amplitude,  $\delta^{(v)}$ is the vertex
correction for either $\bar{e} \nu_e W$ interaction or $\bar{\mu}
\nu_\mu W$ interaction ( since we assume the mass degeneracy for
the first two generations of sleptons, the two
corrections are same), and $\delta^{(b)}$ denotes box diagram
correction.

(1) Vertex corrections

Similar to Eq.(\ref{vertex correction}), the correction to
 $\bar{f}_1 f_2 W$ interaction can be expressed as
\begin{eqnarray}
g_L^{\bar{f}_1 f_2 W} \delta^{(v)}
   &=& \Gamma^{\bar{f}_1 f_2 W}(q^2) - \frac{1}{2} g_L^{\bar{f}_1 f_2 W}
    \biggl \{ \Sigma_{f_1}(m_{f_1}^2) + \Sigma_{f_2}(m_{f_2}^2)
    \biggr\}.
\end{eqnarray}
For the $\bar{e} \nu_e W$ interaction, we have $g_L^{\bar{e} \nu_e
W^-} = - \frac{g}{\sqrt{2}}$,
\small
\begin{eqnarray}
(4 \pi)^2 \Sigma_{e_L}(m_e^2) &=& | g_L^{\bar{\tilde{\chi}}^0_i e
\tilde{e}_L^\ast} |^2 (B_0 + B_1) (m_e^2, m_{\tilde{e}_L},
m_{\tilde{\chi}_i^0} ) + | g_L^{\bar{\tilde{\chi}}_j^- e
\tilde{\nu}_e^\ast } |^2
    (B_0 + B_1) (m_e^2, m_{\tilde{\nu}_e}, m_{\tilde{\chi}_j^-}),
\nonumber  \\
(4 \pi)^2 \Sigma_{\nu_e}(m_{\nu_e}^2) &=& |
g_L^{\bar{\tilde{\chi}}^0_i \nu_e \tilde{\nu}_e^\ast} |^2 (B_0 +
B_1) (m_{\nu_e}^2, m_{\tilde{\nu}_e}, m_{\tilde{\chi}_i^0} ) + |
g_L^{\bar{\tilde{\chi}}_j^+ \nu_e \tilde{e}_L^\ast } |^2
    (B_0 + B_1) (m_{\nu_e}^2, m_{\tilde{e}_L}, m_{\tilde{\chi}_j^+}),
    \nonumber \\
 (4 \pi)^2 \Gamma_{\bar{e} \nu_e W^-} &=&
    - ( g_L^{\bar{\tilde{\chi}}_{i}^0 e \tilde{e}_L^\ast} )^\ast
    g_L^{\bar{\tilde{\chi}}^+_j \nu_e \tilde{e}_L^\ast }
    \nonumber \\
    && \times \left \{
    g_L^{\bar{\tilde{\chi}}_i^0 \tilde{\chi}^+_j  W }  m_{\tilde{\chi}_i^0}
    m_{\tilde{\chi}_j^+}  C_0
    + g_R^{\bar{\tilde{\chi}}_i^0  \tilde{\chi}^+_j  W }
    (-2 C_{24} + \frac{1}{2} ) \right \} (p_{\nu_e}, p_e, m_{\tilde{\chi}_j^+}, m_{\tilde{e}_L}, m_{\tilde{\chi}_i^0}
    )
\nonumber \\
 && - ( g_L^{\bar{\tilde{\chi}}_{j}^- e \tilde{\nu}_e^\ast} )^\ast
    g_L^{\bar{\tilde{\chi}}^0_i \nu_e \tilde{\nu}_e^\ast }
    \nonumber \\
    && \times \left \{
    g_L^{\bar{\tilde{\chi}}_j^- \tilde{\chi}^0_i  W }  m_{\tilde{\chi}_i^0}
    m_{\tilde{\chi}_j^-}  C_0
    + g_R^{\bar{\tilde{\chi}}_j^-  \tilde{\chi}^0_i  W }
    (-2 C_{24} + \frac{1}{2} ) \right \} (p_{\nu_e}, p_e, m_{\tilde{\chi}_i^0}, m_{\tilde{\nu}_e}, m_{\tilde{\chi}_j^-}
    )
\nonumber \\
&& + 2 ( g_L^{\tilde{\chi}_{i}^0 e \tilde{e}_L^\ast } )^\ast
 g_L^{\tilde{\chi}_{i}^0 \nu_e \tilde{\nu}_e^\ast }
 g^{\tilde{e}_L^\ast \tilde{\nu}_e  W} C_{24}(p_{\nu_e}, p_e, m_{\tilde{\nu}_e},
    m_{\tilde{\chi}_{i}^0}, m_{\tilde{e}_L}).
\end{eqnarray}
\normalsize
In the above equations, summation over
 $i=1$ to $5$ $(\tilde{\chi}_i^0)$ and $j=1$ to 2
$(\tilde{\chi}_j^\pm)$ is implied. The coupling $g_L$ takes the following
forms
\begin{eqnarray}
&&g_L^{\bar{\tilde{\chi}}^0_i \nu_e \tilde{\nu}_e^\ast} =
\frac{g}{\sqrt{2}} ( N_{i1}^\ast \tan \theta_W - N_{i2}^\ast );
\quad \quad g_L^{\bar{\tilde{\chi}}^0_i e \tilde{e}_L^\ast} =
\frac{g}{\sqrt{2}}
( N_{i1}^\ast \tan \theta_W + N_{i2}^\ast ); \nonumber \\
&& g_L^{\bar{\tilde{\chi}}^+_j \nu_e \tilde{e}_L^\ast} = - g
U_{j1}^\ast;  \quad  \quad \quad \ \ \quad  \quad \quad \quad \quad
g_L^{\bar{\tilde{\chi}}^-_j e \tilde{\nu}_e^\ast} = - g V_{j1}^\ast \nonumber \\
&& g_L^{\bar{\tilde{\chi}}_i^0 \tilde{\chi}^+_j  W } =
\frac{g}{\sqrt{2}}  ( \sqrt{2} V_{j1}^\ast N_{i2} - V_{j2}^\ast
N_{i3} ); \quad   g_R^{\bar{\tilde{\chi}}_i^0 \tilde{\chi}^+_j W } =
\frac{g}{\sqrt{2}}  ( \sqrt{2} U_{j1} N_{i2}^\ast +
U_{j2} N_{i4}^\ast ); \nonumber \\
&& g_L^{\bar{\tilde{\chi}}_j^- \tilde{\chi}^0_i W } = -
g_R^{\bar{\tilde{\chi}}_i^0 \tilde{\chi}^+_j W };  \quad \quad
g_R^{\bar{\tilde{\chi}}_j^- \tilde{\chi}^0_i W } = -
g_L^{\bar{\tilde{\chi}}_i^0 \tilde{\chi}^+_j W }; \quad \quad
g^{\tilde{e}_L^\ast \tilde{\nu}_e W} = - \frac{g}{\sqrt{2}},
\nonumber
\end{eqnarray}
and for the three-point loop functions, since we take their external
momentum to be zero, their expressions are greatly simplified:
\begin{eqnarray}
C_0 (m_1,m_2,m_3) & =& -\frac{1}{m_3^2} \left \{ - \frac{(1+a)
\ln(1+a)}{a b} + \frac{(1+a+b) \ln(1+a+b)}{(a+b) b} \right \} \nonumber \\
C_{24} (m_1,m_2,m_3) & =& \frac{\Delta}{4} - \frac{1}{4} \ln
\frac{m_3^2}{\mu^2} - \frac{1}{2} \left \{ \frac{-2(1+a)^2
\ln(1+a)}{4 a b} \right . \nonumber \\
&& \left . + \frac{- 3 b (a +b ) + 2 (1+ a+b)^2 \ln (1+a+b)}{4 b
(a+b)} \right \} \nonumber
\end{eqnarray}
with $a = \frac{m_2^2 -m_3^2}{m_3^2} $ and $b = \frac{m_1^2 -
m_2^2}{m_3^2}$.

(2) Box corrections

The box diagram contributions to the $\mu \to \nu_\mu e \bar{\nu}_e$
amplitude can be expressed as
\begin{eqnarray}
i T & = & i \left \{
    M(1) + M(2) + M(3) + M(4) \right \}
    \bar{u}_e \gamma^\mu P_L v_{\nu_e}
    \bar{u}_{\nu_\mu} \gamma^\mu P_L u_{\mu}.
\end{eqnarray}
Taking into account the normalization of the tree-level
amplitude, $- g^2/2 M_W^2$, the box diagram contributions can be
written as
\begin{eqnarray} \delta^{(b)} &=& -\frac{2
M_W^2}{g^2} \sum_{i=1}^4 M(i).
\end{eqnarray}
with each $M(i)$ given by
\begin{eqnarray}
16 \pi^2 M(1) &=& (g_L^{\bar{\tilde{\chi}}_i^0 e
\tilde{e}_L^\ast})^*
        g_L^{\bar{\tilde{\chi}}_i^0 \mu \tilde{\mu}_L^\ast }
       (g_L^{\bar{\tilde{\chi}}^+_j \nu_\mu \tilde{\mu}_L^\ast})^*
        g_L^{\bar{\tilde{\chi}}^+_j \nu_e \tilde{e}_L^\ast}
    D_{27}(m_{\tilde{\mu}_L}, m_{\tilde{e}_L}, m_{\tilde{\chi}^+_j},
    m_{\tilde{\chi}_i^0}),
\nonumber \\
16 \pi^2 M(2) &=& (g_L^{\bar{\tilde{\chi}}_j^- e
\tilde{\nu}_e^\ast})^*
    g_L^{\bar{\tilde{\chi}}_j^- \mu \tilde{\nu}_\mu^\ast }
    (g_L^{\bar{\tilde{\chi}}_i^0 \nu_\mu \tilde{\nu}_\mu^\ast})^*
    g_L^{\bar{\tilde{\chi}}_i^0 \nu_e \tilde{\nu}_e^\ast}
   D_{27}(m_{\tilde{\nu}_\mu}, m_{\tilde{\nu}_e},
    m_{\tilde{\chi}_j^-}, m_{\tilde{\chi}_i^0}), \nonumber
    \\
16 \pi^2 M(3) &=& \frac{1}{2} m_{\tilde{\chi}_i^0}
m_{\tilde{\chi}_j^-}
    g_L^{\bar{\tilde{\chi}}^+_j \nu_e \tilde{e}_L^\ast}
    g_L^{\bar{\tilde{\chi}}_j^- \mu \tilde{\nu}_\mu^\ast }
    (g_L^{\bar{\tilde{\chi}}_i^0 \nu_\mu \tilde{\nu}_\mu^\ast})^*
    (g_L^{\bar{\tilde{\chi}}_i^0 e \tilde{e}_L^\ast})^*
   D_0(m_{\tilde{\nu}_\mu}, m_{\tilde{e}_L}, m_{\tilde{\chi}_j^-},
    m_{\tilde{\chi}_i^0}),
\nonumber \\
16 \pi^2 M(4) &=& \frac{1}{2} m_{\tilde{\chi}_i^0}
m_{\tilde{\chi}_j^-}
    g_L^{\bar{\tilde{\chi}}_i^0 \nu_e \tilde{\nu}_e^\ast}
    g_L^{\bar{\tilde{\chi}}_i^0 \mu \tilde{\mu}_L^\ast}
    (g_L^{\bar{\tilde{\chi}}^+_j \nu_\mu \tilde{\mu}_L^\ast})^*
    (g_L^{\bar{\tilde{\chi}}_j^- e \tilde{\nu}_e^\ast } )^*
   D_0(m_{\tilde{\mu}_L}, m_{\tilde{\nu}_e}, m_{\tilde{\chi}_j^-},
    m_{\tilde{\chi}_i^0}).
\nonumber
\end{eqnarray}
Here all the $D$-functions are evaluated at the zero
momentum-transfer limit. Noting the fact that $m_{\tilde{\mu}_L} \simeq
m_{\tilde{e}_L} \simeq m_{\tilde{\nu}_\mu} \simeq m_{\tilde{\nu}_e}
$, we may write the $D$ functions as
\begin{eqnarray}
D_0 (m_1, m_1, m_2, m_3 )&=& \frac{1}{m_3^4} \left \{
\frac{-(1+a)\ln(1+a)}{ab^2} \right.\nonumber \\
&& \left. + \frac{ -b(a+b) +
((a+b)(1+a+b)+ b)\ln(1+a+b)}{b^2(a+b)^2} \right\}, \nonumber \\
D_{27} (m_1, m_1, m_2, m_3 )&=& - \frac{1}{2 m_3^2} \left \{
\frac{(1+a)^2 \ln(1+a)}{2 a b^2} \right.\nonumber \\
&& \left. - \frac{ (1+a+b) ( - b(a+b) + ((a+b)(1+a )+ b)\ln(1+a+b))
}{2 b^2(a+b)^2} \right\}. \nonumber
\end{eqnarray}

%-----------------------------------------------------------------------------

\end{document}